\documentclass[prc,aps,a4paper,groupedaddress,superscriptaddress,nofootinbib
,preprintnumbers,twocolumn]{revtex4}
\usepackage{graphicx} 
\usepackage{epsfig}
\usepackage[small]{subfigure}
\usepackage{setspace}
\usepackage[T1]{fontenc}
\usepackage[utf8]{inputenc}
\usepackage[english]{babel}
\usepackage{mathrsfs}
\usepackage{braket}
\usepackage[overload]{empheq} 
\usepackage{amssymb}
\usepackage{amsmath}
\usepackage{amsfonts}
\usepackage{siunitx}
\usepackage{amsopn}
\usepackage{mathtools}
\usepackage{float}
\usepackage{chemformula}
\usepackage{comment} 
\usepackage{tikz}
\usepackage{booktabs}
\usepackage{multirow}
\usepackage{verbatim}
\usepackage{xcolor}
\usepackage{ulem}
\usepackage{soul}
\usetikzlibrary{arrows}
\usepackage[colorlinks,linkcolor=blue,urlcolor=blue,citecolor=blue]{hyperref}

\begin{document}

\title{Light clusters in warm magnetized stellar matter: \\ equation of state and thermodynamic response}
\author{Luigi Scurto}
\email{scurto@lns.infn.it}
\affiliation{INFN, Laboratori Nazionali del Sud, I-95123 Catania, Italy}
\author{Stefano Burrello} 
\email{burrello@lns.infn.it} 
\affiliation{INFN, Laboratori Nazionali del Sud, I-95123 Catania, Italy}
\author{Maria Colonna}
\email{colonna@lns.infn.it}
\affiliation{INFN, Laboratori Nazionali del Sud, I-95123 Catania, Italy}

\begin{abstract}
Finite-temperature equations of state (EOSs) with a controlled treatment of composition-dependent effects are becoming increasingly important 
for modeling proto-neutron stars and binary neutron star merger remnants, where warm matter may coexist with strong magnetic fields. At sub-saturation densities, light nuclear clusters may also emerge with sizeable abundances. For beta-equilibrated matter, with or without neutrino trapping, an interplay between magnetic fields and light-cluster formation naturally arises in determining the matter composition: charge neutrality and weak equilibrium transmit the effects of Landau quantization to the baryonic sector, 
modifying the equilibrium charge content; at the same time, light-cluster formation also favors the increase of the proton fraction by binding protons into nuclear clusters.
In this work, we investigate this interplay within a generalized relativistic mean-field framework, in which light clusters up to alpha particles are 
included as explicit degrees of freedom and their in-medium dissolution is described through phenomenological binding-energy shifts.
We show that the formation and subsequent dissolution of light clusters, combined with magnetic-field effects, leave characteristic signatures 
in the matter pressure, the isothermal squared speed of sound, and the heat capacity, leading to significant modifications of the thermodynamic 
stiffness of the EOS and of the heat-storage properties of warm stellar matter. 
Furthermore, we investigate the impact of the isovector terms of the EOS, namely its symmetry energy, on these features. 
These results provide microscopic insights relevant to modeling the hydrodynamic and thermal evolution of proto-neutron stars and neutron star 
merger remnants, while establishing a baseline for the development of more comprehensive finite-temperature EOSs for compact-star  applications.
\end{abstract}

\maketitle

\section{Introduction}
The study of dense nuclear matter has entered a stage in which the equation of state (EOS) is constrained not only by the properties of cold, catalyzed neutron stars (NSs) but also by multimessenger observations of a broader class of compact-star systems~\cite{AbbottPRL2017, MillerAJL2019, RaaijmakersAJL2021, PangNAT2023}. Proto-NSs---including strongly magnetized newborn NSs---and binary-neutron-star merger remnants can contain matter that is hot, lepton-rich, strongly asymmetric and, in some cases, subject to intense magnetic fields~\cite{oertelRMP2017, PonsLRCP2019, PeregoEPJA2019,CiolfiGRG2020,BurgioPPNP2021, SarinGRG2021, EspinoPRL2024, BernuzziMNRAS2025}. With the advent of third-generation gravitational-wave detectors such as the Einstein Telescope~\cite{MaggioreJCAP2020, AbacJCAP2026}, the interpretation of these systems therefore requires EOSs able to describe not only the cold high-density sector, which controls masses, radii and tidal deformabilities~\cite{AnnalaPRL2018,  LattimerARNPS2021}, but also the thermal response of matter with various composition under dynamical conditions~\cite{MiravetPRD2025}. In particular, binary-neutron-star merger
simulations are sensitive to the treatment of the thermal sector of the
EOS, which can affect the dynamics and structure of the remnant, as well
as the post-merger gravitational-wave and kilonova signals~\cite{HammondPRD2021, RaithelPRL2023}.

Among the thermodynamic quantities influenced by the matter composition, the speed of sound and the heat capacity are particularly informative. The adiabatic speed of sound governs the propagation of pressure perturbations and enters the modeling of hydrodynamic modes, post-merger dynamics, and stability against compression~\cite{Kojo2AAPPS021, ScurtoPRD2024}, while its isothermal counterpart provides a complementary diagnostic of the EOS stiffness at fixed temperature. 
The heat capacity determines how efficiently matter stores thermal energy and is therefore a key ingredient for the thermal evolution of hot remnants~\cite{FieldsAJL2023}, proto-NSs and crustal regions~\cite{FortinPRC2010, burrelloPRC2015}, together with neutrino emissivities and transport coefficients~\cite{PotekhinSSR2015, burrelloPRC2016}. Although they are not direct astrophysical observables, these quantities characterize, respectively, the mechanical
response of matter to compression and its thermal response to energy
deposition or cooling, thereby linking the microscopic composition to
the dynamical and thermal evolution of hot compact-star systems~\cite{RaithelPRD2021}.

The scenario is enriched for strongly magnetized systems, where intense magnetic fields can modify the microscopic properties of matter~\cite{MostAJL2025, AdhikariPPNP2026}. Even when the magnetic pressure does not dominate the EOS, the field changes the density of states of charged particles through Landau quantization~\cite{BroderickAJ2000, ScurtoPRC2023}. The leading effect is expected to arise from the lepton sector, because of the much smaller critical magnetic field of the electron~\cite{HardingRPP2006}. Previous relativistic mean field (RMF) studies 
in beta-equilibrated matter 
have shown that this modification can significantly affect the equilibrium electron
fraction at low densities and temperatures~\cite{RabhiPRC2011,
ScurtoPRC2024}. Through charge neutrality and weak equilibrium, the
effect might be transmitted to the total proton content and to the free-neutron abundance, leading to a broader rearrangement of the
baryonic sector and, more generally, of the composition of warm stellar matter~\cite{ScurtoPRC2023, ScurtoPRC2024}.

In this respect, the sub-saturation-density region is of particular relevance~\cite{burrelloEPJA2022, BurrelloPRC2025}. At finite temperature, matter cannot always be described as a gas of interacting nucleons alone~\cite{HorowitzNPA2006, radutaPRC2010}. Light clusters such as deuterons, tritons, helions, and alpha particles may appear with sizable abundances, depending on density, temperature, and isospin asymmetry~\cite{SumiyoshiPRC2008,typelPRC2010,FischerPRC2020, WangPRC2026}. Their formation redistributes baryons between free nucleons and bound states, thereby modifying the chemical composition and potentially affecting the thermal and mechanical response of matter~\cite{burEPJA2022, BurrelloPRC2026}.

Since light-cluster abundances are strongly sensitive to the total
proton fraction~\cite{typelPRC2010,ZhangPRC2017}, the
magnetic-field-induced rearrangement of the charged component generates
a non-trivial interplay between Landau quantization and cluster
formation. 

A consistent treatment of light clusters requires the inclusion of in-medium effects~\cite{paisPRC2019, WangPRC2024}. With increasing density, Pauli blocking and self-energy corrections reduce the cluster binding energies and eventually lead to their dissolution~\cite{ropkeNPA2011}. This physics has been incorporated in quantum-statistical approaches and in generalized mean-field models, where clusters are treated as explicit quasiparticle degrees of freedom embedded in a nucleonic medium~\cite{typelPRC2010, ZhangPRC2017, WangPRC2024}. 

The present work extends these studies of dilute nuclear matter with light-cluster degrees of freedom and in-medium effects~\cite{WangPRC2026, BurrelloPRC2026}, building on recent RMF analysis of the joint effect of temperature and magnetic fields on neutron-star matter~\cite{ScurtoPRC2024}. These two complementary lines of investigation are finally brought together in a generalized relativistic mean-field (gRMF) framework~\cite{typelPRC2010} for warm magnetized nuclear matter with light clusters.
The particle content is restricted to nucleons, electrons, electron neutrinos when relevant, and light clusters up to alpha particles. The magnetic field is included through Landau quantization of charged particles~\cite{ScurtoPRC2023}, while the in-medium dissolution of clusters is described through phenomenological binding-energy shifts~\cite{typelPRC2010}. We consider both neutrino-free beta-equilibrated matter and matter with trapped electron neutrinos at fixed lepton fraction, which represent two limiting
weak-equilibrium regimes that may occur in different regions and evolutionary stages of hot compact objects. Neutrino trapping is
expected in the optically thick interiors of proto-NSs and
in the dense regions of merger remnants,  where neutrino diffusion is slower than the relevant dynamical evolution~\cite{EndrizziPRD2020, EspinoPRL2024}. Conversely,
neutrino-free beta-equilibrium conditions are more appropriate for
neutrino-transparent outer layers, disks and outflows, as well as for
later stages following substantial deleptonization
~\cite{oertelRMP2017, PeregoEPJA2019}. In addition, we explore
the sensitivity to the isovector sector of the EOS through controlled variations of the
density dependence of the \(\rho\)-meson coupling, keeping the isoscalar
terms and the symmetry energy at saturation fixed. In this way, we
investigate the interplay between weak-equilibrium conditions,
magnetic-field effects, light-cluster formation, and the sub-saturation
behavior of the symmetry energy in determining the properties of warm
stellar matter. By comparing calculations with and without light clusters,
and by varying the magnetic field, temperature, weak-equilibrium conditions, and isospin dependence of the EOS, we assess how these ingredients modify the electron fraction and the
free-neutron abundance, as well as the formation, relative hierarchy, and
dissolution of light clusters. We then analyze how the corresponding changes in the composition are reflected in thermodynamic properties
such as the energy per baryon, pressure, fixed-temperature squared speed of sound, and heat capacity per baryon. 

We note that the role of light clusters is also central in intermediate-energy heavy-ion collisions~\cite{onoPPNP2019, WangPRC2023}, which offer a laboratory counterpart to assess the EOS~\cite{zhengPRC2018,tsangNAT2024, ShvedovPRC2025} and in particular
the density dependence of the symmetry energy,~\cite{NatowitzPRL2010, zhengPLB2017, LynchPLB2022}, in a complementary manner to the multimessenger astrophysics~\cite{HuthNAT2022, tsangNAT2024}. Specifically, the cluster features enter the EOS of sub-saturated matter, thus providing crucial experimental information on the low-density behavior of the underlying effective interaction and on the in-medium modification of nuclear correlations~\cite{qinPRL2012, CustodioPRL2025}.

The systematic study of stellar matter undertaken here can therefore provide useful microscopic guidance for future combined constraints of the nuclear EOS from heavy-ion-collision data and astrophysical observations~\cite{HuthNAT2022, tsangNAT2024}. More
generally, it contributes to the development of finite-temperature EOS
models in which composition-dependent effects, light clusters, and
isovector properties are treated consistently.

The paper is organized as follows. In Sec.~\ref{sec:framework} we present the gRMF formalism for magnetized warm matter with light clusters, including the in-medium binding-energy shifts. In Sec.~\ref{sec:results} we discuss the composition and thermodynamic response of neutrino-free and neutrino-trapped matter in the presence of strong magnetic fields. Conclusions and perspectives are given in Sec.~\ref{sec:conclusions}.

\section{Theoretical framework}
\label{sec:framework}
Warm neutron-star matter at total baryon density $n_{b}$ and temperature $T$ is described within a gRMF formalism. Besides nucleons, namely neutrons \(n\) and protons \(p\), light clusters are included as explicit quasiparticle degrees of freedom: deuterons \(d\), tritons \(t\), \(^3\mathrm{He}\) nuclei \(h\), and alpha particles, \(\alpha\). In this formalism, the interaction between baryonic degrees of freedom (nucleons and light clusters) is mediated by three types of mesons: the isoscalar-scalar meson $\sigma$, the isoscalar-vector meson $\omega$ and the isovector-vector meson $\rho$.

The Lagrangian density that describes our system is given by

\begin{equation}
\mathcal{L}
=
\sum_{\kappa=n,p,t,h}\mathcal{L}_{\kappa}
+\mathcal{L}_{d}
+\mathcal{L}_{\alpha}
+\mathcal{L}_{l}
+\mathcal{L}_{\sigma}
+\mathcal{L}_{\omega}
+\mathcal{L}_{\rho}
+\mathcal{L}_{A}
\label{eq:L}
\end{equation}

The fermionic baryonic contributions are
\begin{equation}
\mathcal{L}_\kappa
=
\bar{\psi}_\kappa
\left(
\gamma_\mu iD_\kappa^\mu
-
M_\kappa^*
\right)
\psi_\kappa,
\qquad \kappa \in \{n,p,t,h\} 
\label{eq:L_ferm_baryons}
\end{equation}

The deuteron, treated as a spin-1 boson, is described by
\begin{eqnarray}
\mathcal{L}_{d}&=&\frac{1}{4}\left(iD_d^\mu \phi_d^\nu - iD_d^\nu \phi_d^\mu\right)^*\left(iD_{d\mu}\phi_{d\nu}-iD_{d\nu}\phi_{d\mu}\right)\nonumber\\
&&-\frac{1}{2} M_d^{*2}\,\phi_d^{\mu *}\phi_{d\mu}
\label{eq:L_d}
\end{eqnarray}
while the \(\alpha\) particle, treated as a spin-0 boson, is described by
\begin{equation}
\mathcal{L}_{\alpha}
=
\frac{1}{2}
\left(iD_\alpha^\mu\phi_\alpha\right)^*
\left(iD_{\alpha\mu}\phi_\alpha\right)
-
\frac{1}{2} M_\alpha^{*2}\,\phi_\alpha^*\phi_\alpha.
\label{eq:L_alpha}
\end{equation}

In Eqs.~\eqref{eq:L_ferm_baryons}, \eqref{eq:L_d}, and
\eqref{eq:L_alpha}, the covariant derivative for a generic baryonic
species \(j\in\mathcal{B}\) is defined as
\begin{equation}
iD_j^\mu
=
i\partial^\mu
-
g_{\omega j}\,\omega^\mu
-
g_{\rho j}\,\rho_3^\mu
-
q_j A^\mu ,
\label{eq:covariant_derivative}
\end{equation}
where
\(\mathcal{B}=\{n,p\}\cup\mathcal{C}\), with
\(\mathcal{C}=\{d,t,h,\alpha\}\), $A^\mu$ denotes the electromagnetic four-potential associated
with the external magnetic field, and \(q_j\) is the electric charge
of species \(j\).

The corresponding in-medium quasiparticle effective mass is given by
\begin{equation}
M_j^*
=
A_j M
-
B_j^{\rm eff}
-
g_{\sigma j}\sigma,
\qquad j\in\mathcal{B},
\label{eq:cluster_eff_mass}
\end{equation}
where $A_{j}$ denotes the total mass number and $M$ is the nucleon rest mass. The quantity $B_j^{\rm eff}$ denotes the effective in-medium binding energy, which is nonzero only for clusters; for nucleons one has $B_\tau^{\rm eff}=0$, with $\tau = n, p$. The prescription used to compute the cluster effective binding energies is presented in Sec.~\ref{subsec:binding}.

For the isoscalar
mesons, the cluster-meson couplings, $g_{m j}$, which appear in Eqs.~\eqref{eq:covariant_derivative} and \eqref{eq:cluster_eff_mass}, are taken to scale the nucleon-meson coupling with $A_{j}$,
\begin{equation}
g_{mj}
=
A_j g_{m},
\qquad
m=\sigma,\omega.
\label{eq:cluster_isoscalar_couplings}
\end{equation}
For the isovector \(\rho\) meson, the third component of isospin $I_{3j}$ is
absorbed into the effective coupling,
\begin{equation}
g_{\rho j}
=
I_{3j}g_{\rho}.
\label{eq:cluster_isovector_couplings}
\end{equation}
With this convention, the \(\rho\)-meson contribution vanishes for the isospin-symmetric clusters \(d\) and \(\alpha\). In this work, we will consider density-dependent RMF parametrizations~\cite{typelPRC2010} in which the nucleon-meson couplings are considered to be functions of the total baryonic density $n_{b}$. In particular, the density dependence of the nucleon couplings to the
\(\sigma\) and \(\omega\) mesons is parametrized as
\begin{equation}
g_m(n_b)
=
g_m(n_{\rm sat})\,a_m
\frac{1+b_m(x+d_m)^2}
     {1+c_m(x+d_m)^2},
\quad
m\in\{\sigma,\omega\},
\label{eq:isoscalar_couplings}
\end{equation}
where \(x=n_b/n_{\rm sat}\), with \(n_{\rm sat}\) denoting the saturation
density. The density dependence of the \(\rho\)-meson coupling is instead
given by
\begin{equation}
g_\rho(n_b)
=
g_\rho(n_{\rm sat})
\exp\left[-a_\rho(x-1)\right].
\label{eq:a_rho}
\end{equation}
The quantities \(g_m(n_{\rm sat})\) and \(g_\rho(n_{\rm sat})\)
denote the corresponding meson--nucleon couplings at saturation density,
whereas the coefficients \(a_\rho\), \(a_m\), \(b_m\), \(c_m\), and \(d_{m}\) characterize their density
dependence and satisfy the constraints
specified in Ref.~\cite{typelPRC2010}.

The leptonic term in the Lagrangian density is given by 
\begin{eqnarray}
\mathcal{L}_{l} =  
\bar{\psi}_e
\left[
\gamma_\mu\left(i\partial^\mu+e A^\mu\right)-m_e
\right]  \psi_e  
+ \bar\psi_{\nu_e} i\gamma_\mu \partial^\mu \psi_{\nu_e},
\label{eq:L_leptons}
\end{eqnarray}
where leptons (electrons, $e$, and electron neutrinos, $\nu_{e}$) are treated as a free Fermi gas and where $m_{e}$ is the electron mass.
The electromagnetic-field contribution to the Lagrangian density is
given by the standard Maxwell term
\begin{equation}
\mathcal{L}_{A}
=
-\frac{1}{4}F_{\mu\nu}F^{\mu\nu},
\qquad
F_{\mu\nu}=\partial_\mu A_\nu-\partial_\nu A_\mu .
\label{eq:L_em}
\end{equation}

Finally, the mesonic contributions in the Lagrangian density are given by 
\begin{equation}
\mathcal{L}_{\sigma}
=
\frac{1}{2}
\left(
\partial_\mu \sigma \,\partial^\mu \sigma
-
m_\sigma^2 \sigma^2
\right),
\label{eq:L_sigma}
\end{equation}
\begin{equation}
\mathcal{L}_{\omega}
=
-\frac{1}{4} G_{\mu\nu}G^{\mu\nu}
+\frac{1}{2} m_\omega^2 \omega_\mu \omega^\mu,
\label{eq:L_omega}
\end{equation}
\begin{equation}
\mathcal{L}_{\rho}
=
-\frac{1}{4} \mathbf{H}_{\mu\nu}\!\cdot\!\mathbf{H}^{\mu\nu}
+\frac{1}{2} m_\rho^2 \,\boldsymbol{\rho}_\mu\!\cdot\!\boldsymbol{\rho}^{\,\mu},
\label{eq:L_rho}
\end{equation}
with
\begin{equation}
G_{\mu\nu}=\partial_\mu \omega_\nu-\partial_\nu \omega_\mu,
\qquad
\mathbf{H}_{\mu\nu}=\partial_\mu \boldsymbol{\rho}_\nu-\partial_\nu \boldsymbol{\rho}_\mu
\label{eq:field_tensors}
\end{equation}
and $m_\sigma$, $m_\omega$ and $m_\rho$ the masses of the three mesons.

\subsection{In medium binding energies} 
\label{subsec:binding}
The effective binding energies that enter the Dirac mass of light clusters in Eq.~\eqref{eq:cluster_eff_mass} are given by 
\begin{equation}
    B_j^{\rm eff}=B_j^{\rm vac}-\Delta B_j 
\end{equation}
where $B_j^{\rm vac}$ is the binding energy in the vacuum and, following the generalized RMF prescription of Ref.~\cite{typelPRC2010},
the in-medium reduction of the cluster binding energy is written as
\begin{equation}
\Delta B_j(n_b,Y_p,T)
=
\tilde n_j
\left(
1+\frac{\tilde n_j}{2\tilde n_j^0(T)}
\right)\delta B_j(T),
\label{eq:DeltaB_Typel_RMF}
\end{equation}
where
\begin{equation}
\tilde n_j
=
\frac{2n_b}{A_j}\left[Z_jY_p+N_j(1-Y_p)\right],
\label{eq:ntilde_short}
\end{equation}
\(Z_j\) and \(N_j\) denote the proton and neutron numbers, respectively, $Y_{p}$ is the total proton fraction, and
\begin{equation}
\tilde n_j^0(T)=\frac{B_j^{\rm vac}}{\delta B_j(T)}.
\label{eq:ntilde0_short}
\end{equation}
The quantity \(\delta B_j(T)\) is obtained from the zero-density limit of
the Pauli-blocking shift. For the deuteron, one has
\begin{align}
\delta B_d(T) = \frac{a_{d,1}}{T^{3/2}}
\Bigg[
\frac{1}{\sqrt{y_d}}
-
\sqrt{\pi}\,a_{d,3}\,
\exp\!\left(x_{d}^2\right)
\operatorname{erfc}\!\left(x_{d}\right)
\Bigg],
\label{eq:deltaB_d_short}
\end{align}
with
\begin{equation}
y_d=1+\frac{a_{d,2}}{T}, \qquad x_{d} = a_{d, 3} \sqrt{y_{d}}
\label{eq:y_d_short}
\end{equation}
whereas for \(c=t,h,\alpha\),
\begin{equation}
\delta B_c(T)
=
\frac{a_{c,1}}{T^{3/2}}
\frac{1}{y_c^{3/2}},
\qquad
y_c=1+\frac{a_{c,2}}{T}.
\label{eq:deltaB_tha_short}
\end{equation}
\begin{table}[tbp!]
\centering
\caption{Parameters entering the temperature-dependent functions
\(\delta B_j(T)\) in the generalized RMF prescription of Ref.~\cite{typelPRC2010}.}
\label{tab:Typel_short}
\begin{tabular}{c c c c}
\hline\hline
Cluster \(j\) &
\(a_{j,1}\) [MeV$^{5/2}$ fm$^3$] &
\(a_{j,2}\) [MeV] &
\(a_{j,3}\) \\
\hline
\(d\)      & 38386.4 & 22.5204 & 0.2223 \\
\(t\)      & 69516.2 & 7.49232 & -- \\
\(h\)      & 58442.5 & 6.07718 & -- \\
\(\alpha\) & 164371  & 10.6701 & -- \\
\hline\hline
\end{tabular}
\end{table}
The values used for the parameters in Eqs.~\eqref{eq:deltaB_d_short},~\eqref{eq:y_d_short} and~\eqref{eq:deltaB_tha_short} are taken from Ref.~\cite{typelPRC2010} and are shown in Tab.~\ref{tab:Typel_short}. This parametrization ensures that \(\Delta B_j\to 0\) as \(n_b\to 0\),
so that the vacuum limit is recovered. 

\subsection{Distribution functions and thermodynamic quantities} 
\label{Sec_thermo}
At finite temperature, the density and thermodynamic 
quantities are evaluated
from the quasiparticle occupation probabilities, described by
Fermi--Dirac and Bose--Einstein distributions,
for fermions and bosons, respectively.  For both cases, the particle and antiparticle occupation functions are written in the compact form
\begin{equation}
f_j^{\pm}(k)
=
\left[
\exp\left(\frac{E_j(k)\mp\mu_j^*}{T}\right)
+\eta_j
\right]^{-1},
\label{eq:distribution_neutral}
\end{equation}
where the upper sign refers to particles and the lower sign to
antiparticles, with \(\eta_j=+1\) for fermions and \(\eta_j=-1\) for
bosons.\footnote{In practice, the antiparticle contribution of bosonic clusters is negligible in the thermodynamic conditions explored here, but it is kept in the formal expressions for completeness.}
In Eq.~\eqref{eq:distribution_neutral}, the quasiparticle energies are defined as 
\begin{equation}
E_j(k)=\sqrt{k^2+M_j^{*2}},
\label{eq:neutral_dispersion}
\end{equation}
while $\mu_j^*$ is the particle effective chemical potential, given by
\begin{equation}
\mu_j^*
=
\mu_j
-
g_{\omega j}\omega_0
-
g_{\rho j}\rho_{03}
-
\Sigma_j^{(r)},
\label{eq:mu_eff}
\end{equation}
where the rearrangement contribution is split as
\begin{equation}
\Sigma_j^{(r)}=\Sigma_{g}^{(r)}+\Sigma_{B,j}^{(r)}.
\label{eq:Sigma_r_split}
\end{equation}
The term 
\begin{align}
\Sigma_{g}^{(r)}=
\sum_{j\in \mathcal{B}}
\Bigg(
\frac{\partial g_{\omega j}}{\partial n_b}\,\omega_0\,n_j^{(v)}
+
\frac{\partial g_{\rho j}}{\partial n_b}\,\rho_{03}\,n_j^{(v)} - 
\frac{\partial g_{\sigma j}}{\partial n_b}\,\sigma\,n_j^{(s)}
\Bigg)
\label{eq:Sigma_g_r}
\end{align}
is due to the density dependence of the
meson couplings. Since the in-medium cluster binding-energy shifts considered in this work (Eq.~\eqref{eq:DeltaB_Typel_RMF}) depend explicitly on density, an additional
mass-shift rearrangement term must be included,
\begin{equation}
\Sigma_{B,j}^{(r)}
=
\sum_{\kappa\in\mathcal C}
n_\kappa^{(s)}
\frac{\partial \Delta B_\kappa}{\partial n_j^{(v)}}.
\label{eq:Sigma_B_r}
\end{equation}

The Fermi-Dirac distribution functions for leptons are obtained from Eq.~\eqref{eq:distribution_neutral} replacing the baryonic effective mass and effective chemical potential with the corresponding lepton mass and chemical potential.

For a generic species $j$, the vector density $n_j^{(v)}$, and the scalar density $n_j^{(s)}$,
entering Eqs.~\eqref{eq:Sigma_g_r} and~\eqref{eq:Sigma_B_r}, are given by: 
\begin{eqnarray}
n_j^{(v)}
&=&
\frac{\gamma_j}{2\pi^2}
\int_0^\infty dk\,k^2
\left[
f_j^+(k)-f_j^-(k)
\right]; 
\label{eq:nv_neutral} \\
n_j^{(s)}
&=&
\frac{\gamma_j}{2\pi^2}
\int_0^\infty dk\,k^2
\frac{M_j^*}{E_j(k)}
\left(
f_j^++f_j^-
\right),
\label{eq:ns_neutral}
\end{eqnarray}
where $\gamma_j=2S_{j}+1$ is the spin multiplicity.

The total energy density and pressure of the system are
\begin{equation}
\mathcal{E}
=
\sum_{j}\mathcal{E}_j^{\rm kin}
+
\frac{1}{2}m_\sigma^2\sigma^2
+
\frac{1}{2}m_\omega^2\omega_0^2
+
\frac{1}{2}m_\rho^2\rho_{03}^2,
\label{eq:energy_density_total}
\end{equation}
and 
\begin{eqnarray}
P
= && 
\sum_{j}P_j^{\rm kin}
-
\frac{1}{2}m_\sigma^2\sigma^2
+
\frac{1}{2}m_\omega^2\omega_0^2
+
\frac{1}{2}m_\rho^2\rho_{03}^2
\nonumber \\ 
+ &&
P_g^{(r)}
+
P_b^{(r)},
\label{eq:pressure_total}
\end{eqnarray}
where the sum runs over all the considered quasiparticles.
The kinetic contributions to the energy density and pressure  are given by
\begin{equation}
\mathcal{E}_j^{\rm kin}
=
\frac{\gamma_j}{2\pi^2}
\int_0^\infty dk\,k^2 E_j(k)\left(f_j^++f_j^-\right),
\label{eq:eps_neutral_fermion}
\end{equation}
and
\begin{equation}
P_j^{\rm kin}
=
\frac{\gamma_j}{6\pi^2}
\int_0^\infty dk\,\frac{k^4}{E_j(k)}
\left(f_j^++f_j^-\right).
\label{eq:press_neutral_fermion}
\end{equation}

The rearrangement terms in Eq.~\eqref{eq:pressure_total} are given by 
\begin{equation}
P_g^{(r)}=n_b\,\Sigma_{g}^{(r)},
\label{eq:Pg_r}
\end{equation}
and
\begin{equation}
P_b^{(r)}=
\sum_{j\in \mathcal{B}} n_j^{(v)}\,\Sigma_{B,j}^{(r)}.
\label{eq:PB_r}
\end{equation}

The entropy density follows from the thermodynamic relation
\begin{equation}
\mathcal{S} =
\frac{\mathcal{E}+ P
-\sum_j \mu_j n_j^{(v)}}{T},
\label{eq:entropy_density}
\end{equation}
where the sum runs over all particle species present in the system.

\subsection{Introduction of the Landau quantization scheme}
The thermodynamic expressions introduced in the previous subsection are written in terms of standard three-dimensional momentum integrals. However, in the presence of a uniform magnetic field, the energy spectra and phase-space measures of charged particles must be replaced by their Landau-quantized forms, while neutral particles retain their
standard expressions.

The coupling of charged particles to the external magnetic field is introduced through the covariant derivative, Eq.~\eqref{eq:covariant_derivative}.
We thus consider a uniform external magnetic field of fixed strength directed along the \(z\)-axis, \(\mathbf{B}=B\hat z\), and adopt the Landau gauge \(A^\mu=(0,0,Bx,0)\).
In the present work, anomalous magnetic moments of particles are neglected, and thus only charged species couple to the magnetic field. Moreover, the charged light clusters are treated as
quasiparticles whose center-of-mass motion is Landau quantized, while possible additional magnetic
couplings associated with their internal structure are neglected.

For a charged species \(j\), the quasiparticle energy depends on the
Landau-level (LL) index \(\lambda\) and on the momentum component \(k_z\)
parallel to the magnetic field,
\begin{equation}
E_{j,\lambda}(k_z)
=
\sqrt{k_z^2+\mathcal{M}_j^2+2\lambda |q_j|B},
\label{eq:charged_dispersion}
\end{equation}
where \(\mathcal{M}_j=M_j^*\) for baryonic quasiparticles and \(\mathcal{M}_j=m_e\) for electrons.

The distribution functions for charged particles are thus given by
\begin{equation}
f_{j,\lambda}^{\pm}(k_z)
=
\left[\exp\left(\frac{E_{j,\lambda}(k_z)\mp\mu_j^*}{T}\right) + \eta_{j}
\right]^{-1}.
\label{eq:distributions_charged}
\end{equation}
Accordingly, for charged particles the phase-space integral is replaced by
\begin{equation}
\int \frac{d^3k}{(2\pi)^3}
\;\longrightarrow\;
\frac{|q_j|B}{2\pi^2}
\sum_{\lambda}\gamma_{j,\lambda}
\int_0^\infty dk_z.
\label{eq:Landau_replacement}
\end{equation}

For the charged spin-\(\tfrac12\) particles, namely \(p\), \(t\), \(h\), and \(e\), one has \(\lambda=0,1,2,\ldots\) and
the degeneracy of the
\(\lambda\)-th LL is
$\gamma_{j,\lambda}=2-\delta_{\lambda 0}$.

For charged bosonic
clusters, for which spin-magnetic splitting is neglected, the same notation
is used with \(\lambda=\ell+1/2\), \(\ell=0,1,2,\ldots\), and with spin
multiplicity \(\gamma_{j,\lambda}=2S_j+1\).

The vector density finally becomes
\begin{equation}
n_j^{(v)}
=
\frac{|q_j|B}{2\pi^2}
\sum_{\lambda=0}^{\infty}\gamma_{j,\lambda}
\int_0^\infty dk_z
\left(f_{j,\lambda}^+ - f_{j,\lambda}^-\right),
\label{eq:nv_charged}
\end{equation}
and all the other quantities introduced in Sec.~\ref{Sec_thermo} are modified accordingly by using the Landau-quantized spectrum and the replacement rule in Eq.~\eqref{eq:Landau_replacement}.

Since we focus on the matter contribution to the EOS, the Maxwell term
\(B^2/2\) is not included in the energy density and pressure. 

It is worth noting that, in the present work, \(P\) denotes the thermodynamic matter pressure obtained
from the grand-canonical thermodynamic potential at fixed external magnetic field. It should not be confused with the mechanical pressure components derived from the energy-momentum tensor, which are generally anisotropic in a magnetized medium~\cite{MostAJL2025}. Magnetization effects on the mechanical stresses are indeed
beyond the scope of the present analysis.

\subsection{Charge neutrality, chemical- and weak-equilibrium conditions}

In the present work, matter is assumed to be in chemical equilibrium and electrically neutral. Charge neutrality requires
\begin{equation}
n_p^{(v)}
+
\sum_{j\in\mathcal{C}} Z_j n_j^{(v)}
=
n_e ,
\label{eq:charge_neutrality}
\end{equation}
so that the total proton fraction is given by
\begin{equation}
Y_p
=
\frac{
n_p^{(v)}+\sum_{j\in\mathcal{C}} Z_j n_j^{(v)}
}{n_b}.
\label{eq:proton_fraction}
\end{equation}
Chemical equilibrium between clusters and nucleons is imposed through
\begin{equation}
\mu_j = N_j\mu_n + Z_j\mu_p,
\qquad
j\in\mathcal{C}.
\label{eq:cluster_chemical_equilibrium}
\end{equation}

We consider both neutrino-free and neutrino-trapped weak-equilibrium
conditions. For neutrino-free matter, neutrinos escape from the system
and \(\mu_{\nu_e}=0\), so that weak equilibrium requires
\begin{equation}
\mu_n-\mu_p=\mu_e .
\label{eq:beta_equilibrium_nufree}
\end{equation}
We note that, at finite temperature, the standard neutrino-free
condition of Eq.~\eqref{eq:beta_equilibrium_nufree} does not necessarily
coincide with the condition obtained by requiring the neutron-decay and
electron-capture rates to balance in neutrino-transparent matter. In
particular, Ref.~\cite{AlfordPRC2018} showed that this effect can be
described by an additional isospin chemical potential, leading to a
moderate change in the proton fraction under the conditions explored in
that work. However, a consistent implementation of such a correction in
the present case would require the calculation of weak-interaction rates
in sub-saturation, clustered, and magnetized matter, which is beyond the
scope of the present study. We therefore retain the standard
chemical-equilibrium condition in
Eq.~\eqref{eq:beta_equilibrium_nufree}.

When electron neutrinos are trapped, \(\mu_{\nu_e}\neq 0\), and the
\(\beta\)-equilibrium condition reads
\begin{equation}
\mu_n-\mu_p=\mu_e-\mu_{\nu_e},
\label{eq:beta_equilibrium_trapped}
\end{equation}
while the total lepton fraction is fixed according to
\begin{equation}
Y_l
= Y_{e} + Y_{\nu_e} = 
\frac{n_e+n_{\nu_e}}{n_{b}},
\label{eq:lepton_fraction}
\end{equation}
where $Y_{e} = n_{e}/n_{b}$ and $Y_{\nu_e} = n_{\nu_e}/n_{b}$ denote the electron and electron-neutrino fractions, respectively.
Finally, we consider two quantities characterizing the thermodynamic
response of the system: the fixed-temperature squared speed of sound and the heat-capacity
per baryon. At finite temperature, the definition of the speed of sound
depends on the thermodynamic path along which the derivative is taken. The hydrodynamic
sound speed is usually evaluated along an adiabatic
trajectory, possibly assuming a frozen composition depending on the relevant
time scales. In the present work, we instead consider the fixed-temperature
quantity
\begin{equation}
c_{s,T}^2
=
\left.
\frac{\partial P}{\partial \mathcal{E}}
\right|_{T,\mathrm{eq}}
=
\frac{
\left.\partial P/\partial n_b\right|_{T,\mathrm{eq}}
}{
\left.\partial \mathcal{E}/\partial n_b\right|_{T,\mathrm{eq}}
},
\label{eq:sound}
\end{equation}
where the derivative is evaluated along the considered equilibrium
sequence, at fixed temperature, imposing charge neutrality and the
appropriate weak-equilibrium condition. Therefore, \(c_{s,T}^2\) should be
interpreted as a diagnostic of the density dependence of the EOS at fixed
temperature.

The equilibrium heat capacity per baryon at constant volume is computed as
\begin{equation}
c_V = \dfrac{1}{n_b}
\left. \dfrac{\partial \mathcal{E}}{\partial T} \right|_{n_b,\mathrm{eq}},
\label{eq:heat}
\end{equation}
where the derivative is taken at fixed total baryon density, with the composition allowed to readjust according to the same chemical- and weak-
equilibrium conditions.

\section{Results}
\label{sec:results}
We first discuss, in
Sec.~\ref{Sub_chem}, the interplay between strong magnetic fields and light-cluster formation in shaping the overall composition of warm \(\beta\)-equilibrated matter, considering
both neutrino-free and neutrino-trapped weak-equilibrium conditions.
Building on this analysis, Sec.~\ref{Sub_EoS} investigates the corresponding effects on the EOS and selected
thermodynamic derivatives.

The DD2 parametrization of Ref.~\cite{typelPRC2010} is adopted as the
reference RMF parametrization in the numerical applications discussed in
Secs.~\ref{Sub_chem} and~\ref{Sub_EoS}. Finally, in
Sec.~\ref{Sub_mod_comp}, we investigate the sensitivity of our results to
controlled variations of the isovector channel of the interaction, which modify the
sub-saturation behavior of the symmetry energy. 

In the following, the magnetic-field strength is expressed in terms of
the dimensionless quantity \(B^*\), defined as
\(B^*=B/B_e^c\), where
\(B_e^c=4.414\times10^{13}\,\mathrm{G}\) is the electron critical
magnetic field, at which the electron cyclotron energy becomes equal to
its rest mass~\cite{HardingRPP2006,ScurtoPRC2024}.
Unless otherwise stated, we consider three temperatures,
\(T=7,10,\) and \(15\) MeV, and four magnetic-field strengths,
\(B^*=0\), \(10^3\), \(3\times10^3\), and \(10^4\). These values span
a range of thermal and magnetic conditions relevant to warm magnetized
stellar matter~\cite{RabhiPRC2011,ScurtoPRC2024}. We focus on the
low-density regime where light-cluster formation is expected to be most
relevant~\cite{typelPRC2010,WangPRC2026,BurrelloPRC2026}. 
Finally, for neutrino-trapped matter, we fix the
total lepton fraction to \(Y_l=0.4\), representative of lepton-rich
stellar matter~\cite{RabhiPRC2011, oertelRMP2017}.

The present treatment is thus restricted to homogeneous matter containing
nucleons and light clusters up to \(\alpha\) particles. Heavy nuclei,
non-homogeneous pasta configurations, and continuum correlations are not included. In particular, continuum correlations in the deuteron channel may partly offset the bound state contribution~\cite{burEPJA2022}. Moreover, since anomalous magnetic moments and possible magnetic modifications of the internal structure and binding energies of the clusters are neglected, the results obtained at the largest field strength considered here should be mainly regarded as an exploratory estimate of the effects associated with the Landau quantization.

\subsection{Composition of magnetized warm matter}
\label{Sub_chem}

\begin{figure} [tbp!]
    \centering    \includegraphics[width=\linewidth,angle=0]{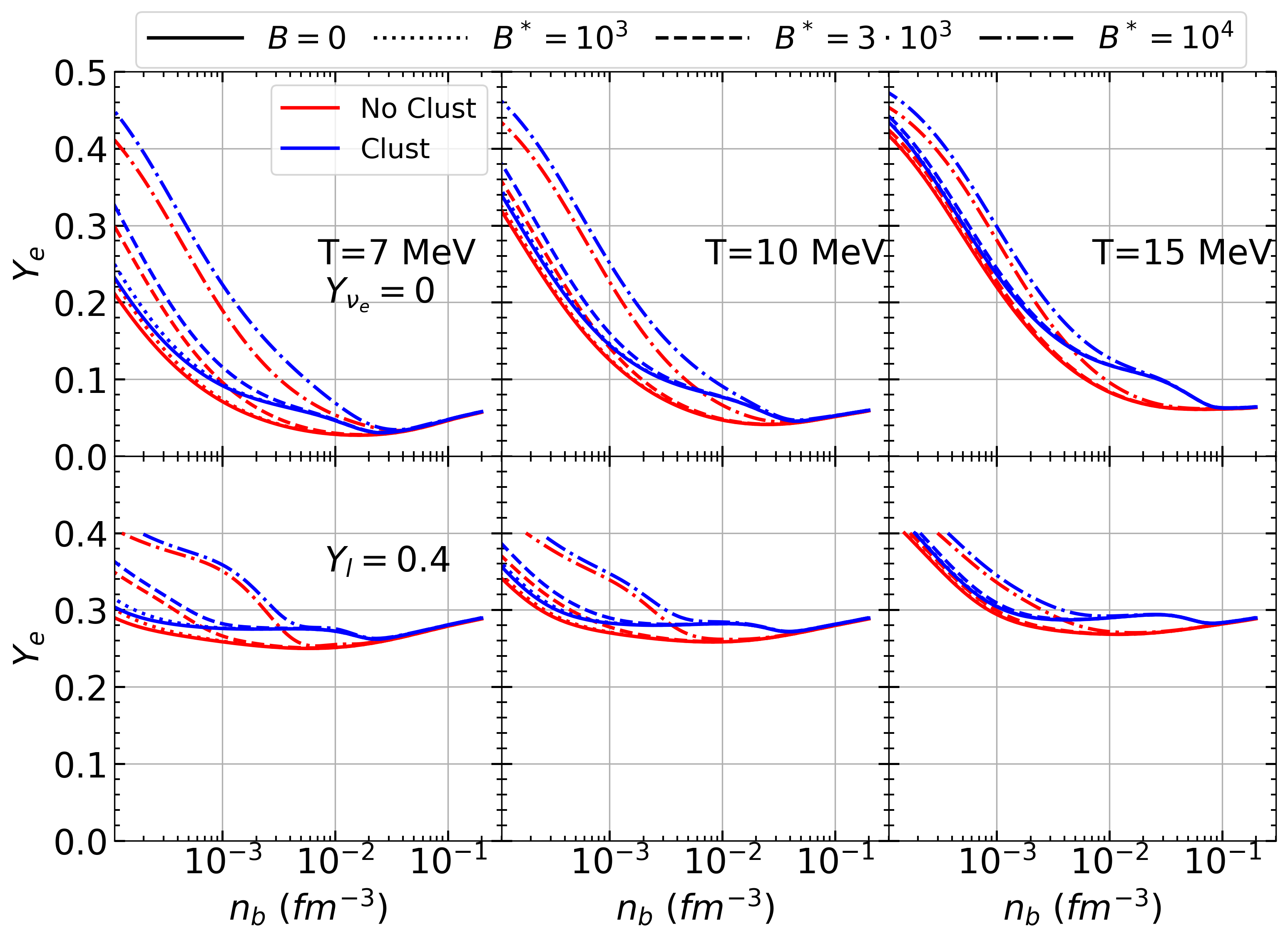}   
    \caption{Electron fraction, $Y_{e}$, in \(\beta\)-equilibrated matter as a function of the total baryon density $n_{b}$ for three temperatures: \(T=7\) MeV (left), \(T=10\) MeV (middle), and \(T=15\) MeV (right). Results are shown for four values of the magnetic field: \(B^*=0\) (solid lines), \(B^*=10^3\) (dotted lines), \(B^*=3\times10^3\) (dashed lines), and \(B^*=10^4\) (dash-dotted lines). The top row corresponds to neutrino-free matter, while the bottom row refers to neutrino-trapped matter with fixed lepton fraction \(Y_l=0.4\). Results obtained with and without light clusters are shown in blue and red, respectively.}
    \label{Fig_ye}
\end{figure}
We first discuss the impact of strong magnetic fields and light clusters
on the composition of low-density and \(\beta\)-equilibrated matter, at finite temperature.
We start from the electron fraction, $Y_{e}$, which, because of charge
neutrality, determines the total proton content of the system. 

Figure~\ref{Fig_ye} shows the electron fraction as a function of the total
baryon density for the thermodynamic conditions specified above. Results obtained  with (blue curves) and without (red curves) light clusters are compared, and for both neutrino-free matter (top panels) and neutrino-trapped matter (bottom panels). It is observed that the magnetic field strongly increases the electron fraction and, consequently, the total proton content, through charge neutrality. The effect is more pronounced at low densities and temperatures, consistently with Ref.~\cite{ScurtoPRC2024}, whereas the enhancement becomes progressively weaker as density and temperature increase, since Landau quantization effects are gradually washed out. This effect is mainly driven by the electron sector, since the strength of Landau
quantization scales as 
$|q_j|/M_j^2$; because of their small mass, electrons are much more strongly affected than protons and light clusters. The size and density dependence of this increase depend, however, on the weak-equilibrium conditions. In neutrino-free matter, the electron fraction can respond directly to the Landau quantization of the electron phase space, producing a steep low-density rise. In neutrino-trapped matter, the electron fraction is already larger because the fixed lepton fraction favors a more proton-rich composition. The magnetic field still produces a sizeable increase in \(Y_e\), although its response is partly constrained by the
fixed lepton fraction: an increase in \(Y_e\) must be accompanied by a reduction in \(Y_{\nu_e}\), thereby smoothing the magnetic-field-induced variation
of \(Y_e\) with density.

At the same time, Fig.~\ref{Fig_ye} shows that, even at $B=0$, light-cluster formation feeds back on the equilibrium charge content of the system, increasing the total proton fraction and, through charge neutrality, the electron fraction. This effect acts in the same direction as the magnetic-field-induced increase but remains modest at the lowest densities. It becomes progressively more pronounced with increasing density and reaches its maximum shortly before the blue and red curves merge, near the density at which light clusters are expected to dissolve.

\begin{figure}
    \centering       \includegraphics[width=\linewidth,angle=0]{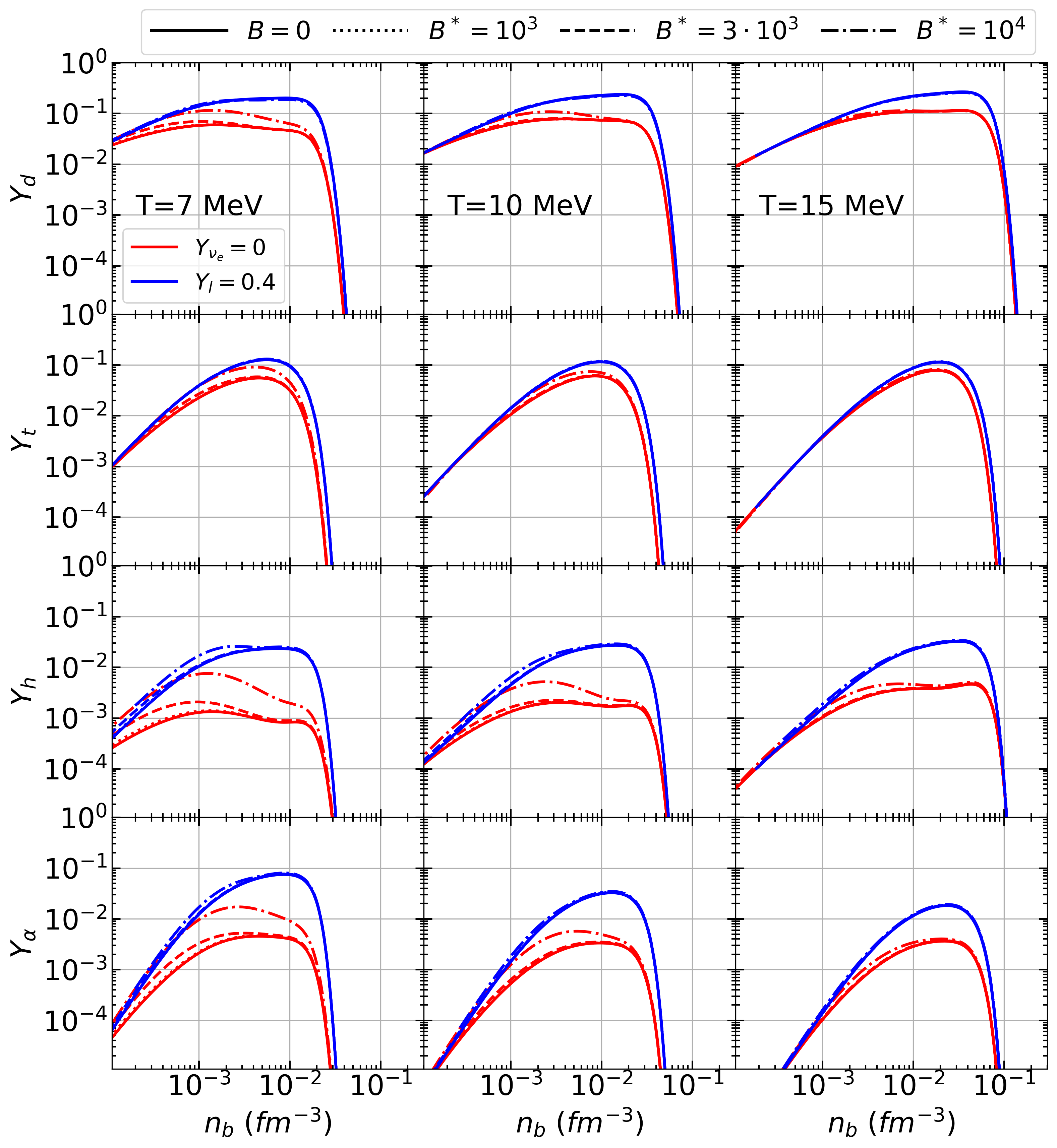}   
    \caption{Light-cluster fractions, $Y_{c}$, with $c \in \{d,t,h,\alpha\}$, in \(\beta\)-equilibrated matter as a function of the total baryon density $n_{b}$ for three temperatures: \(T=7\) MeV (left), \(T=10\) MeV (middle), and \(T=15\) MeV (right). Results are shown for four values of the magnetic field: \(B^*=0\) (solid lines), \(B^*=10^3\) (dotted lines), \(B^*=3\times10^3\) (dashed lines), and \(B^*=10^4\) (dash-dotted lines). Neutrino-free matter is shown in red, while neutrino-trapped matter with fixed lepton fraction \(Y_l=0.4\) is shown in blue. }
    \label{Fig_yc}
\end{figure}
The increase of the total proton content provides also the mechanism behind the enhancement of light-cluster abundances, $Y_{c} = A_c\, n_c/n_b$, with $c \in \{d,t,h,\alpha\}$, observed in Fig.~\ref{Fig_yc}. The cluster fractions are significantly larger in neutrino-trapped matter, particularly for the two species with $Z=2$, namely helions and $\alpha$ particles. This behavior reflects the larger proton fraction associated, through charge neutrality, with the higher lepton content. Indeed, as shown in Ref.~\cite{ZhangPRC2017}, light-cluster abundances are strongly sensitive to the total proton fraction. 

The magnetic field further enhances the cluster fractions. This effect is more pronounced for species with a larger electric charge and, at fixed $Z$, is slightly stronger for clusters with a smaller neutron number. The magnetic-field-induced increase is nevertheless less pronounced in neutrino-trapped matter, where the larger cluster abundances already favored by the fixed lepton fraction leave less room for an additional rearrangement of the composition. Moreover, the enhancement gradually weakens with increasing density and temperature, as the differences between the quantization schemes become progressively less relevant.

\begin{figure}
    \centering
       \includegraphics[width=\linewidth,angle=0]{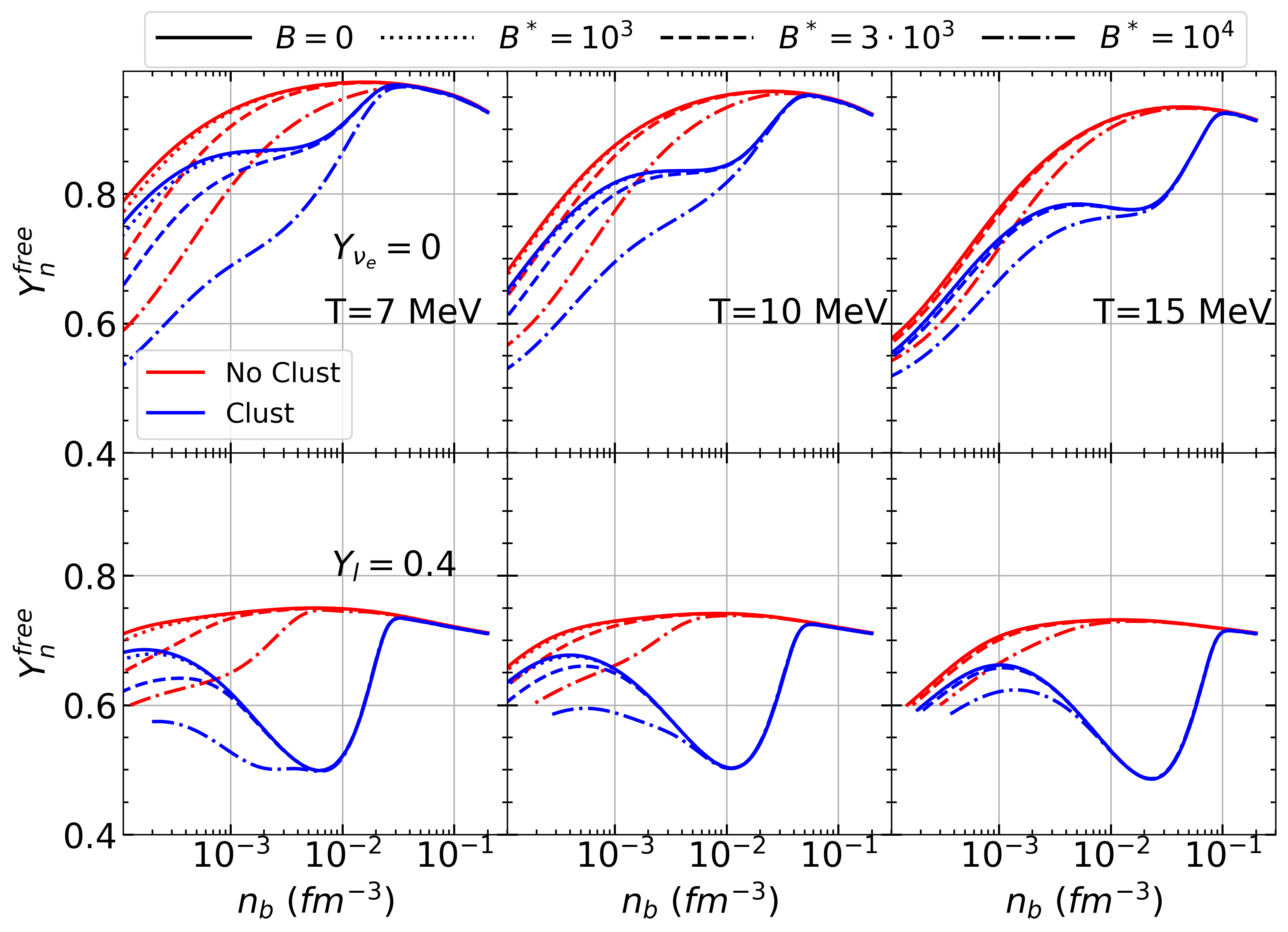}   
    \caption{Free-neutron fraction, $Y_{n}^{\rm free}$,  in \(\beta\)-equilibrated matter as a function of the total baryon density $n_{b}$ for three temperatures: \(T=7\) MeV (left), \(T=10\) MeV (middle), and \(T=15\) MeV (right). Results are shown for four values of the magnetic field: \(B^*=0\) (solid lines), \(B^*=10^3\) (dotted lines), \(B^*=3\times10^3\) (dashed lines), and \(B^*=10^4\) (dash-dotted lines). The top row corresponds to neutrino-free matter, while the bottom row refers to neutrino-trapped matter with fixed lepton fraction \(Y_l=0.4\). Results obtained with and without light clusters are shown in blue and red, respectively.}
    \label{Fig_yn}
\end{figure}
The changes in the proton and cluster fractions are reflected in
Fig.~\ref{Fig_yn}, which displays the free-neutron fraction $Y_{n}^{\rm free} = n_n^{(v)}/n_{b}$ as a function
of the total baryon density.  At low densities, the magnetic field reduces $Y_n^{\rm free}$ by increasing the charge content of the system. At higher densities, light-cluster formation leads to a pronounced depletion, with the free-neutron fraction decreasing to approximately $50\%$ in the density region where clusters are most abundant. This reduction has a twofold origin: light-cluster formation increases the total proton fraction, as shown in Fig.~\ref{Fig_ye}, while also binding part of the neutron population into clusters.

The free-neutron fraction plays a key role in determining the superfluid properties of the neutron-star inner crust~\cite{ChamelLRR2008,FortinPRC2010,PotekhinAA2013,burrelloPRC2015}. This provides further motivation for extending the present study to lower temperatures, where neutron superfluidity and its interplay with light-cluster formation and magnetic fields should be treated consistently, together with the emergence of heavier nuclei and pasta phases~\cite{ScurtoPRC2023}. More generally,
the abundances of free neutrons, protons, and light clusters determine
the relative populations of the stellar medium undergoing
weak-interaction processes and may therefore influence neutrino opacities and charged-current reaction rates. A quantitative
assessment of these effects would, however, require a consistent calculation of the relevant cross sections and in-medium response functions, which lies beyond the scope of the present study
~\cite{ArconesPRC2008, HorowitzPRC2012, FischerPRC2020,
OertelPRC2020, SuleimanPRC2023}.

\subsection{Equation of state and thermodynamic response}
\label{Sub_EoS}

We now focus on the EOS and on selected thermodynamic response
functions of warm \(\beta\)-equilibrated matter, discussing the combined
role of magnetic fields, light-cluster formation, and weak-equilibrium
conditions. 

\begin{figure}[tbp!]
    \centering
\includegraphics[width=\linewidth,angle=0]{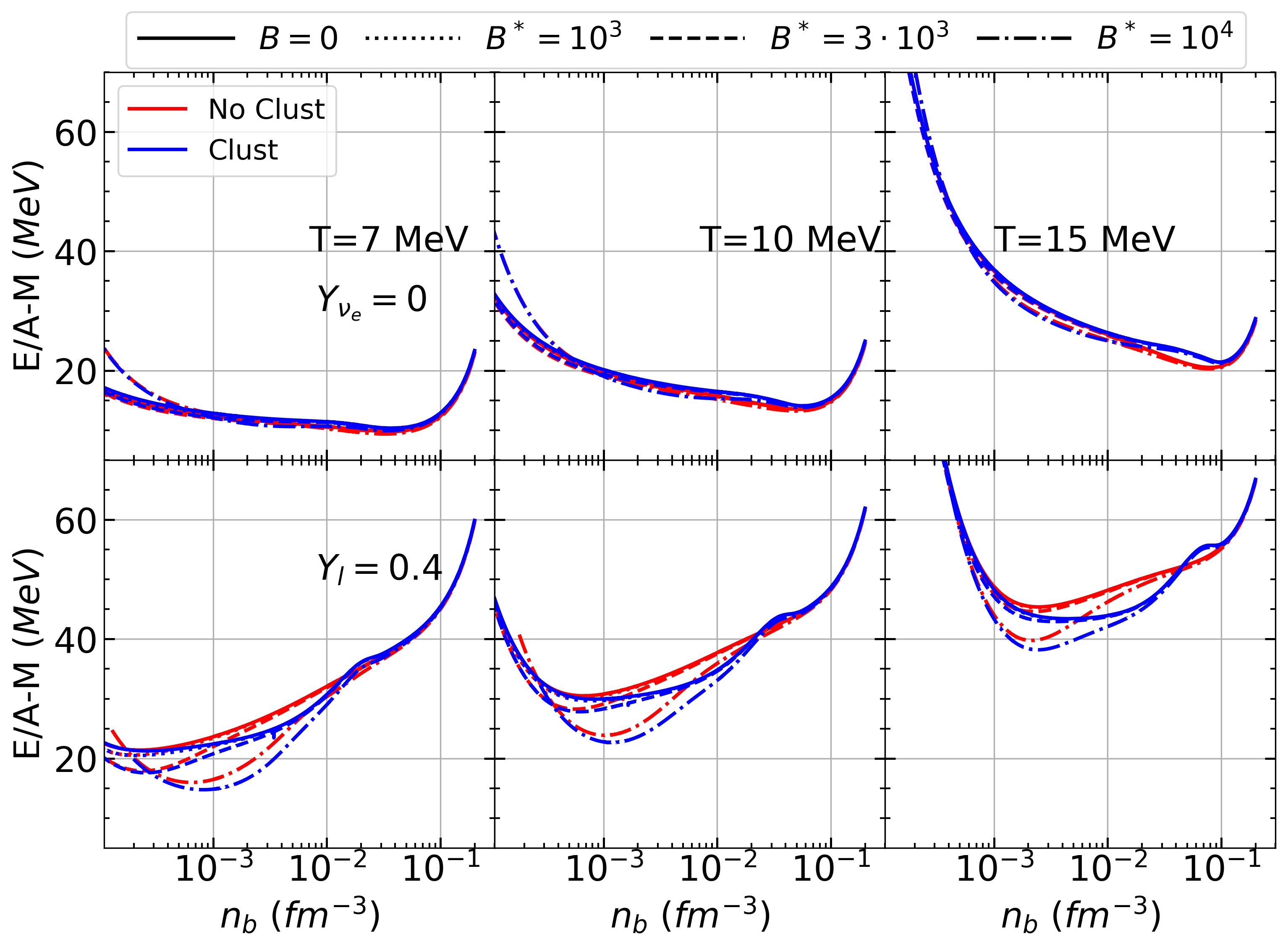}  
\caption{Energy per baryon, $E/A - M$, in \(\beta\)-equilibrated matter as a function
of the total baryon density $n_{b}$ for three temperatures: \(T=7\) MeV (left),
\(T=10\) MeV (middle), and \(T=15\) MeV (right). Results are shown for
four values of the magnetic field: \(B^*=0\) (solid lines),
\(B^*=10^3\) (dotted lines), \(B^*=3\times10^3\) (dashed lines), and
\(B^*=10^4\) (dash-dotted lines). The top row corresponds to
neutrino-free matter, while the bottom row refers to neutrino-trapped
matter with fixed lepton fraction \(Y_l=0.4\). Results obtained with and
without light clusters are shown in blue and red, respectively.}
    \label{Fig_Eb}
\end{figure}

In Fig.~\ref{Fig_Eb}, we show the energy per baryon, $E/A - M$, as a function of the total baryon density for the same temperatures, magnetic-field strengths,
and weak-equilibrium conditions considered in the previous subsection.
In general, the magnetic field lowers the energy per baryon, with the reduction being particularly pronounced around $n_{b}\simeq 10^{-3} \mathrm{fm}^{-3}$, especially in neutrino-trapped matter. An exception occurs at very low densities for the largest magnetic-field strength considered, for which the energy per baryon increases relative to the zero-field result. These trends reflect the interplay between the Landau-quantized spectra of charged particles, the magnetic-field dependence of the level degeneracy, and the resulting rearrangement of the equilibrium composition, consistently with Ref.~\cite{ScurtoPRC2024}.

The effect of light clusters is also non-trivial. Their inclusion lowers the energy per baryon at low densities, while producing a small upward deviation from the cluster-free result in the density region where the
cluster fractions approach their maxima. This feature, which is associated with
the progressive reduction of the in-medium cluster binding energies and the corresponding rearrangement of the equilibrium composition, is more pronounced in neutrino-trapped matter, where cluster abundances are
larger. Temperature affects the magnetic-field and cluster contributions in
different ways. Increasing the temperature reduces the magnetic-field
effect and confines it to a narrower low-density region, because the thermal population of several states smooths the impact of Landau
quantization. By contrast, the density region where light clusters are
relevant is mainly shifted toward larger densities. This modifies the
interplay between magnetic fields and cluster formation, as well as their
relative importance in the EOS.

\begin{figure}[tbp!]
    \centering       \includegraphics[width=\linewidth,angle=0]{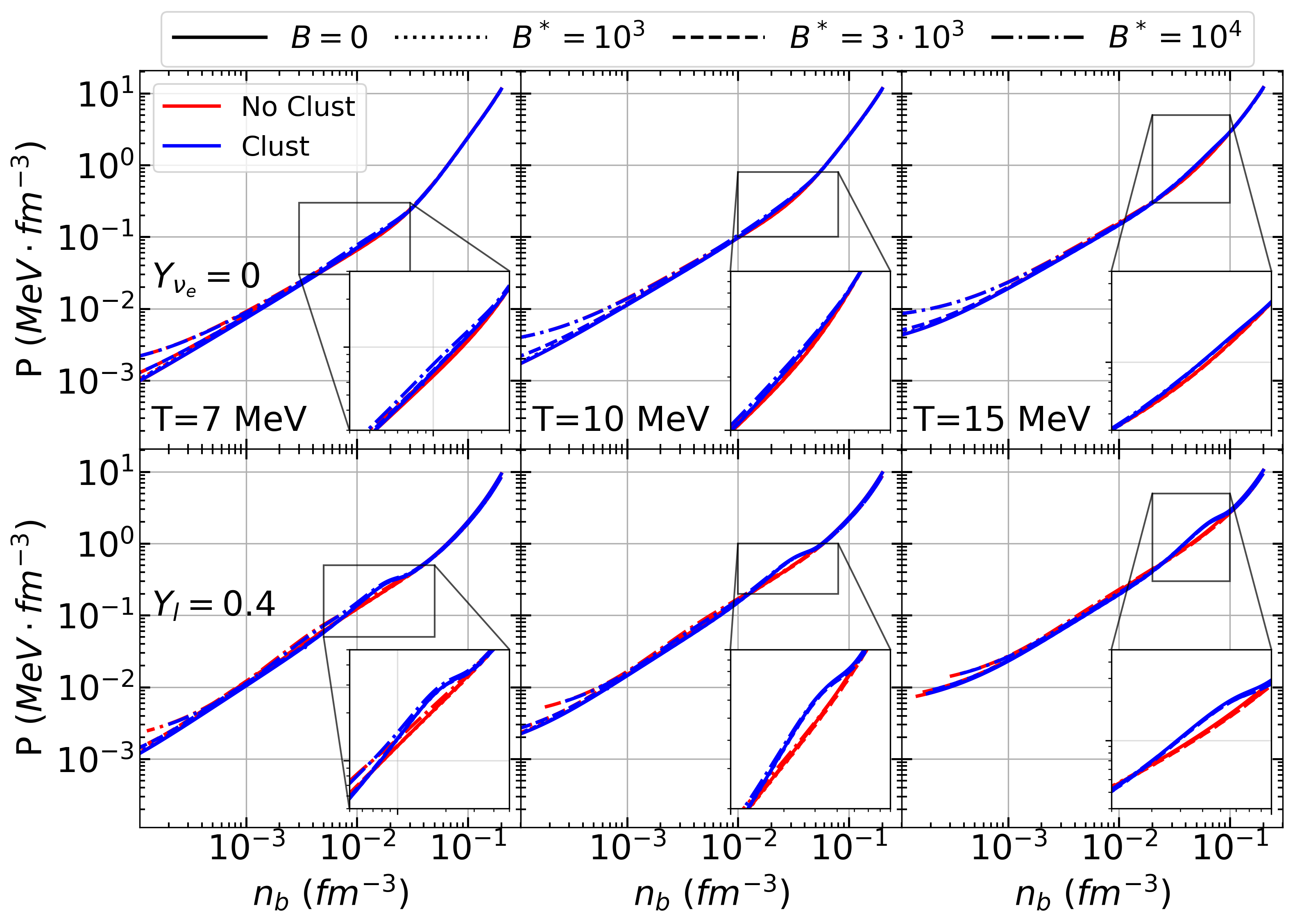}   
\caption{Pressure, $P$, in \(\beta\)-equilibrated matter as a function of the
total baryon density $n_{b}$ for three temperatures: \(T=7\) MeV (left),
\(T=10\) MeV (middle), and \(T=15\) MeV (right). Results are shown for
four values of the magnetic field: \(B^*=0\) (solid lines),
\(B^*=10^3\) (dotted lines), \(B^*=3\times10^3\) (dashed lines), and
\(B^*=10^4\) (dash-dotted lines). The top row corresponds to
neutrino-free matter, while the bottom row refers to neutrino-trapped
matter with fixed lepton fraction \(Y_l=0.4\). Results obtained with and
without light clusters are shown in blue and red, respectively.}
    \label{Fig_P}
\end{figure}

A related signature of cluster formation is visible in Fig.~\ref{Fig_P}, where the pressure, $P$, is shown as a function of the total baryon density. Around the density region where cluster abundances become large, the inclusion of light clusters
leads to a stiffening of the EOS, followed by a softening when clusters
start to dissolve.
By contrast, the magnetic field primarily affects the low-density regime, where Landau quantization modifies the charged-particle phase space and the equilibrium composition. However, its impact on the pressure becomes appreciable only for the largest magnetic-field strength considered.

\begin{figure}[tbp!]
    \centering       \includegraphics[width=\linewidth,angle=0]{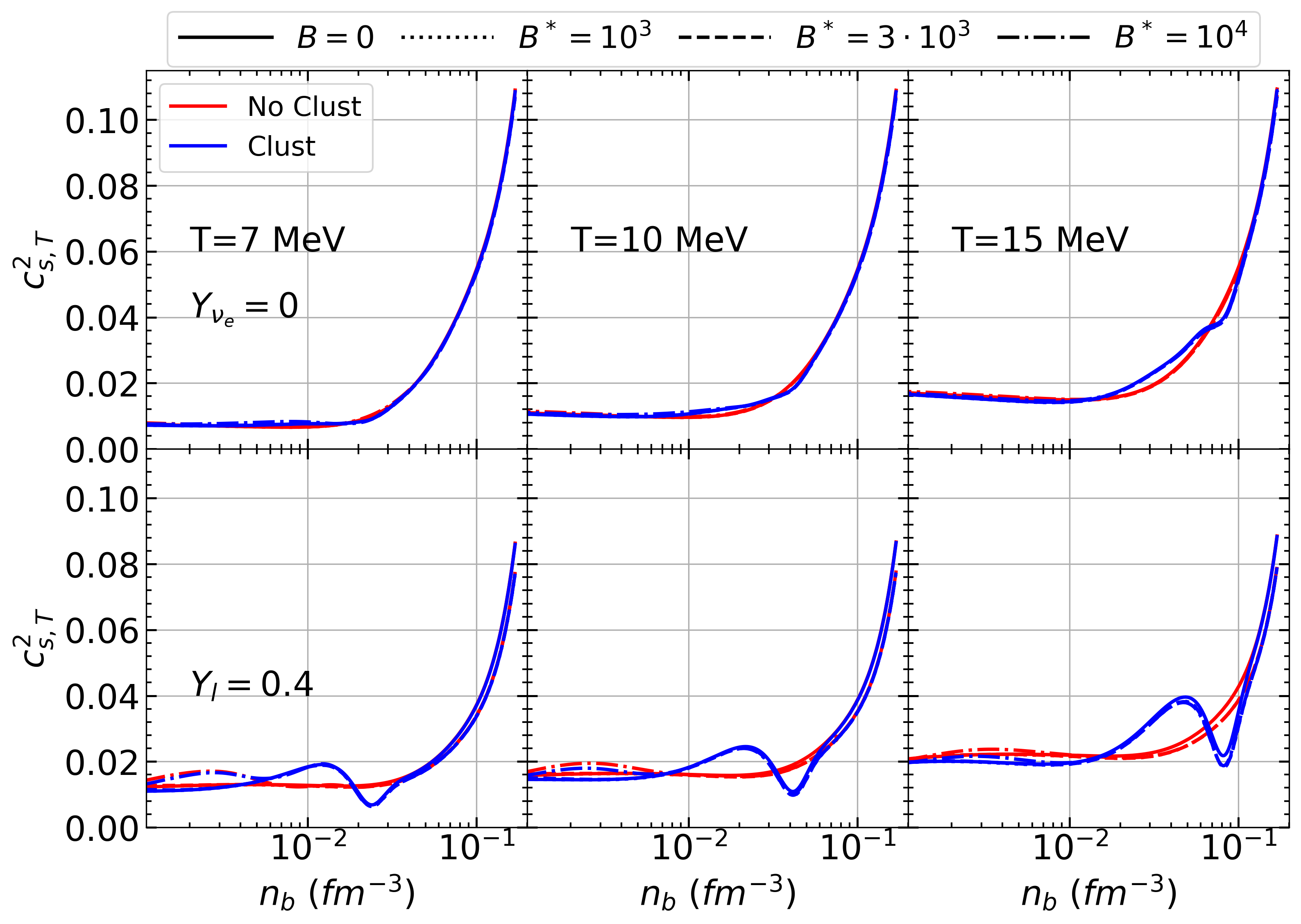}   
\caption{Isothermal squared speed of sound, \(c_{s, T}^2\), in \(\beta\)-equilibrated
matter as a function of the total baryon density for three temperatures:
\(T=7\) MeV (left), \(T=10\) MeV (middle), and \(T=15\) MeV (right).
Results are shown for four values of the magnetic field:
\(B^*=0\) (solid lines), \(B^*=10^3\) (dotted lines),
\(B^*=3\times10^3\) (dashed lines), and \(B^*=10^4\)
(dash-dotted lines). The top row corresponds to neutrino-free matter,
while the bottom row refers to neutrino-trapped matter with fixed lepton
fraction \(Y_l=0.4\). Results obtained with and without light clusters are
shown in blue and red, respectively.}
    \label{Fig_cs}
\end{figure}
The change in the stiffness of the EOS leaves a visible imprint on the isothermal squared speed of sound, \(c_{s, T}^2\), defined in Eq.~\eqref{eq:sound} and shown in Fig.~\ref{Fig_cs}.
The cluster signature is more pronounced in
neutrino-trapped matter, where \(c_{s, T}^2\) develops a clear local maximum
and minimum in the density regions associated, respectively, with the rise
of cluster abundances and their subsequent dissolution. Similar structures are
reminiscent of some features associated with continuous transitions or with the rapid disappearance of different
kinds of two-body correlations, such as pairing or short-range correlations~\cite{burPRC2014,burEPJA2022}. Despite this non-monotonic behavior, \(c_{s,T}^2\) remains positive over the entire density range
considered, indicating that the homogeneous equilibrium branch does not develop an isothermal mechanical instability along the thermodynamic path explored here, as a further check of the consistency of our model.

In neutrino-free
matter, where cluster fractions are smaller, the effect is less prominent
but still visible as a change of slope in the same density interval. The
magnetic-field dependence of \(c_{s, T}^2\) remains mostly restricted to the
lowest densities, consistently with its impact on the pressure and on the
equilibrium composition.

\begin{figure}
    \centering
\includegraphics[width=\linewidth,angle=0]{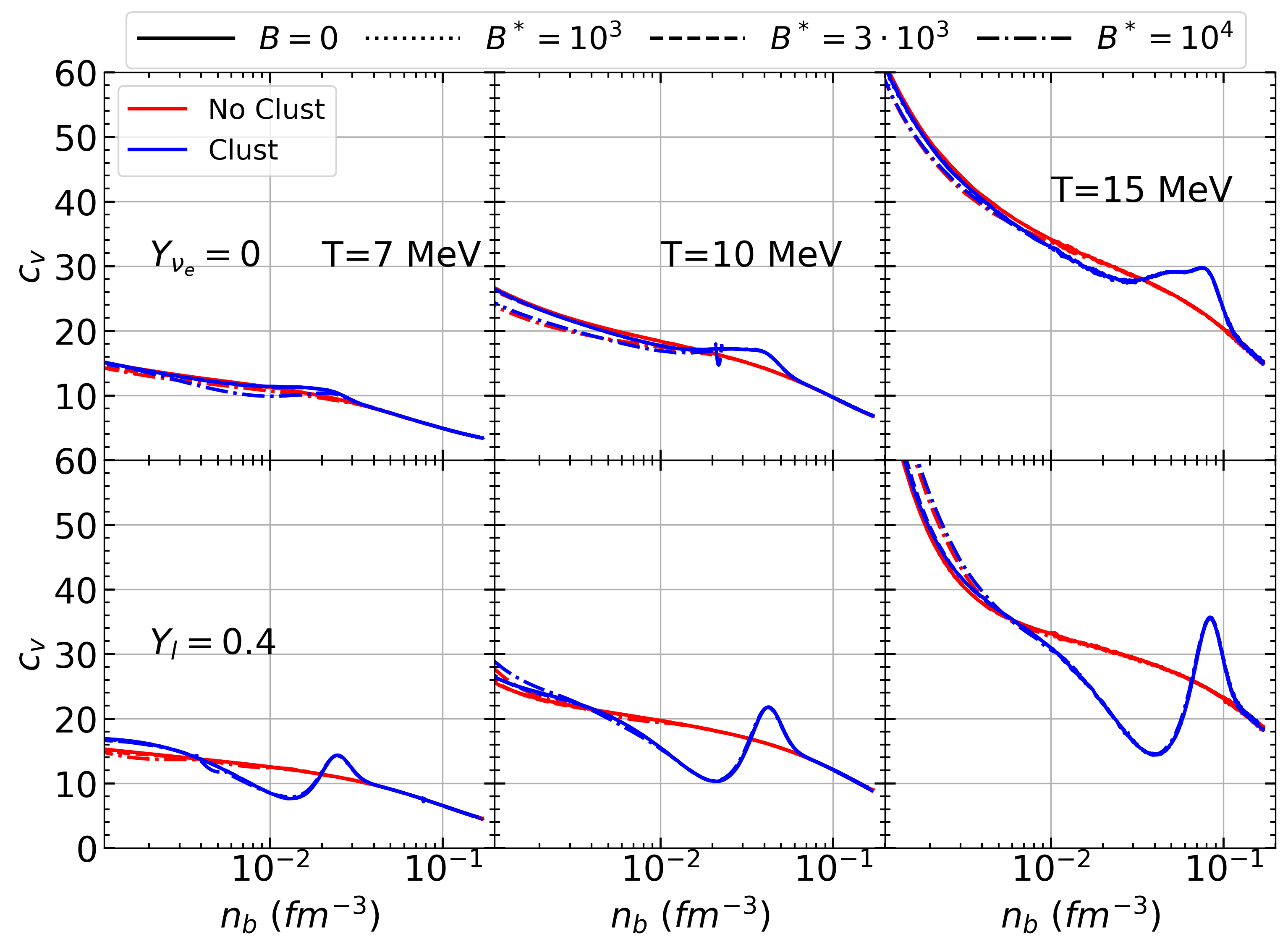}   
    \caption{Heat capacity per baryon, $c_{V}$, in \(\beta\)-equilibrated
matter as a function of the total baryon density for three temperatures:
\(T=7\) MeV (left), \(T=10\) MeV (middle), and \(T=15\) MeV (right).
Results are shown for four values of the magnetic field:
\(B^*=0\) (solid lines), \(B^*=10^3\) (dotted lines),
\(B^*=3\times10^3\) (dashed lines), and \(B^*=10^4\)
(dash-dotted lines). The top row corresponds to neutrino-free matter,
while the bottom row refers to neutrino-trapped matter with fixed lepton
fraction \(Y_l=0.4\). Results obtained with and without light clusters are
shown in blue and red, respectively.}
    \label{Fig_cv}
\end{figure}
Finally, we investigate the heat capacity per baryon, $c_{V}$ (Eq.~\eqref{eq:heat}), of warm \(\beta\)-equilibrated
matter. Figure~\ref{Fig_cv} shows the heat capacity per baryon as a function of the total baryon density. Similarly to the speed of sound, the heat capacity develops a characteristic structure in the density region where cluster formation is most relevant: an initial depletion is followed by a local enhancement near the maximum of the cluster fractions. While the magnetic-field dependence of this observable remains rather modest, the contribution associated with light clusters is sizeable. This structure becomes more pronounced as the temperature increases and the density region affected by cluster formation shifts toward higher values. These results highlight the important role that light clusters may play
in both the mechanical response and the thermal evolution of warm
stellar matter.

\subsection{Role of the symmetry energy}
\label{Sub_mod_comp}
\begin{figure}[tbp!]
\centering    \includegraphics[width=\linewidth,angle=0]{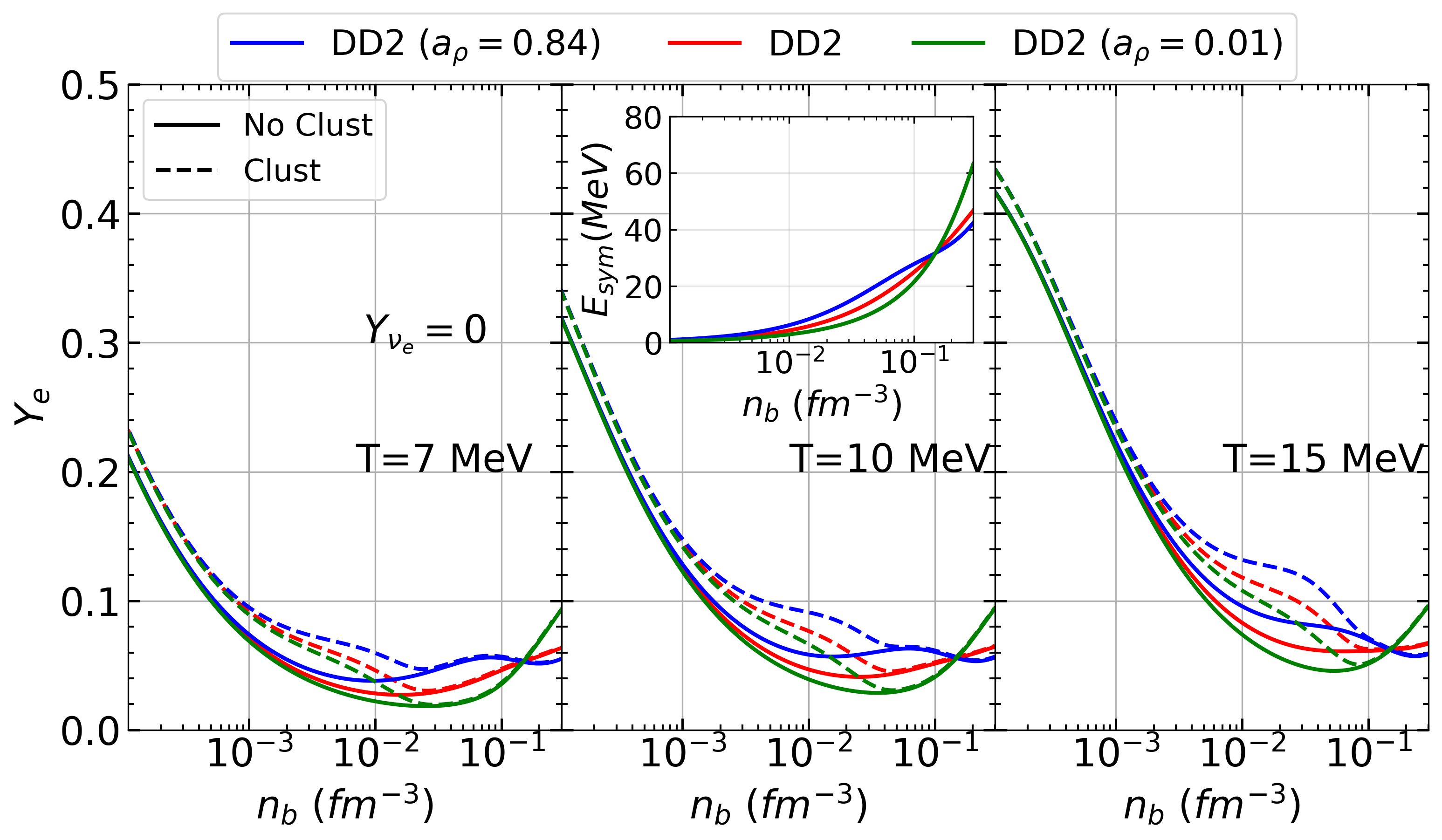}   
    \caption{Electron fraction, $Y_{e}$, for $\beta$-equilibrated matter without neutrinos for three different values of temperature (T=7 MeV on the left, T=10 MeV at the center and T=15 MeV on the right). We compare the results obtained with (dashed lines) and without (solid lines) the inclusion of light clusters, for three DD2-based parametrizations characterized
by \(a_\rho=0.84\) (blue), \(0.52\) (standard DD2, red), and \(0.01\) (green);
the corresponding symmetry-energy curves are displayed in the inset.}
    \label{Fig_ye_comp}
\end{figure}
In the previous subsections, we have shown that light-cluster abundances, and hence their impact on the EOS and the thermodynamic response of matter, are strongly correlated with the total proton fraction. In $\beta$-equilibrated matter, the latter is largely governed by the density dependence of the symmetry energy, $E_{\rm sym}$~\cite{DetovarPRD2021}. We therefore conclude our analysis by investigating how variations in the isovector sector of the interaction affect the equilibrium composition and the resulting cluster abundances and, in turn, the EOS and the thermodynamic quantities of our interest.

For simplicity, we restrict this comparison to neutrino-free matter at $B=0$. This choice allows us to isolate the role of the symmetry energy, avoiding the additional rearrangement of the composition induced by neutrino trapping and Landau quantization.

Following a strategy similar to that adopted in Ref.~\cite{LiAJ2023}, we construct a family of RMF parametrizations derived from the DD2 model employed throughout the previous subsections. All parametrizations share the same isoscalar properties and the same value of the symmetry energy at saturation, $E_{\rm sym}(n_{\rm sat})$, while differing in its density dependence. More specifically, we keep $g_\rho(n_{\rm sat})$ fixed and vary the  $a_\rho$ parameter entering Eq.~\eqref{eq:a_rho}, leaving all other model parameters unchanged. In addition to the standard DD2 parametrization, characterized by
\(a_\rho=0.52\) and a symmetry-energy slope parameter at saturation
density \(L_{\rm sym}=55\) MeV, we consider two modified parametrizations with $a_\rho=0.84$ and $0.01$, corresponding to the extreme lower and upper values of $L_{\rm sym}=30$ and $94$ MeV, respectively~\cite{burrelloPRC2021, BurrelloPRC2025}.
\begin{figure}[tbp!]
    \centering
\includegraphics[width=\linewidth,angle=0]{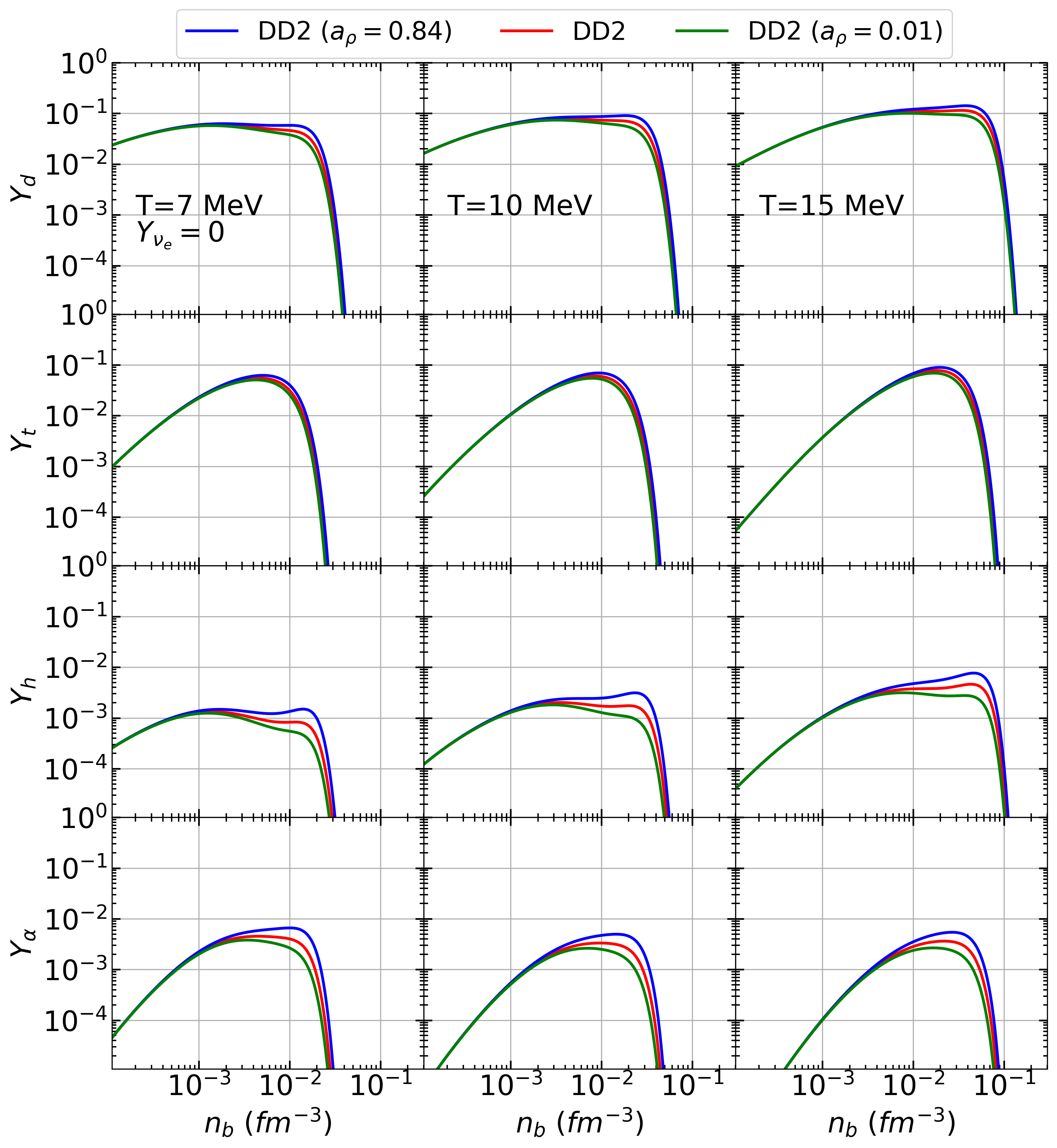}   
\caption{Light-cluster fractions, $Y_c$, with $c=d,t,h,\alpha$, in neutrino-free $\beta$-equilibrated matter at $T=7$ MeV (left), $T=10$ MeV (middle), and $T=15$ MeV (right). Results are shown for three DD2-based parametrizations characterized by $a_\rho=0.84$ (blue), $0.52$ (red), and $0.01$ (green).}
    \label{Fig_yc_comp}
\end{figure}
Since the models differ only in the value of $a_\rho$, this construction allows us to isolate, within the density range considered here, the effects associated with the subsaturation behavior of the symmetry energy. The corresponding  symmetry-energy curves are displayed in the inset of Fig.~\ref{Fig_ye_comp} as functions of the total baryon density. By construction, the three curves cross at saturation density, where $E_{\rm sym}$ is kept unchanged. Below saturation, a smaller (larger) value of $L_{\rm sym}$ corresponds to a larger (smaller) symmetry energy, whereas the ordering is reversed above saturation.

We begin by comparing the electron fractions predicted by the three parametrizations in Fig.~\ref{Fig_ye_comp}. Because charge neutrality is imposed, $Y_{e}$ also represents the total proton fraction. We note that all curves cross at saturation density, as expected, irrespective of whether light clusters are included (dashed lines) or not (solid lines). Below saturation, their ordering follows that of $E_{\rm sym}$: a larger symmetry energy favors a larger proton fraction, thereby reducing the neutron–proton asymmetry of the system. The inclusion of light clusters slightly enhances the spread among the predictions of the different parametrizations, as can be seen
by comparing the separation among the curves obtained with clusters with that among the corresponding cluster-free results. At the same time, the cluster-induced modification of the proton fraction becomes more pronounced as the subsaturation symmetry energy increases. Increasing the temperature further enhances this effect and shifts it toward higher densities, consistently with the corresponding displacement of the cluster-rich region.
\begin{figure}[tbp!]
    \centering
\includegraphics[width=\linewidth,angle=0]{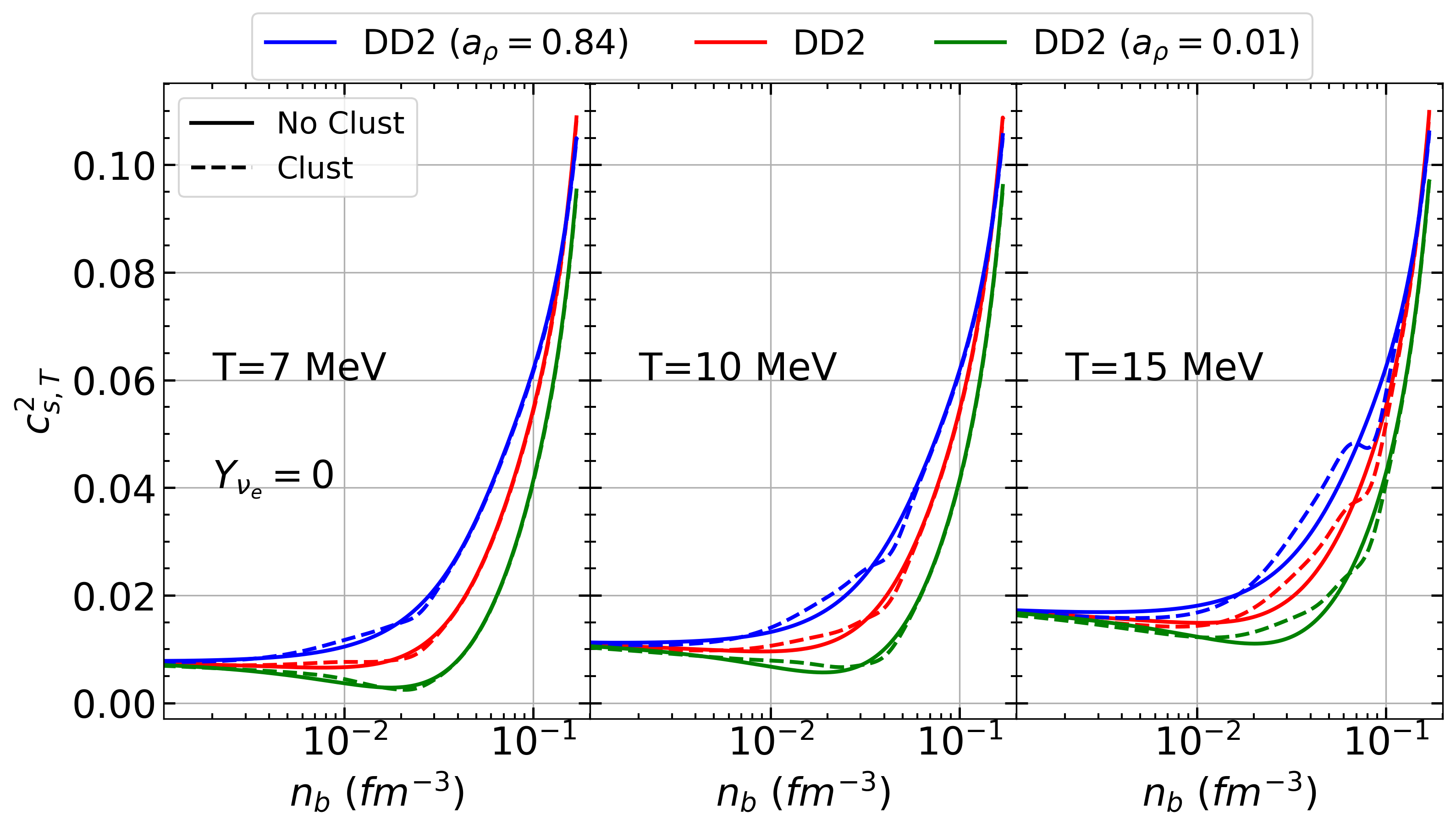}   
\caption{Isothermal squared speed of sound, $c_{s, T}^2$, in neutrino-free $\beta$-equilibrated matter at $T=7$ MeV (left), $T=10$ MeV (middle), and $T=15$ MeV (right). Results obtained with (dashed lines) and without (solid lines) light clusters are shown for three DD2-based parametrizations characterized by $a_\rho=0.84$ (blue), $0.52$ (red), and $0.01$ (green).}
    \label{Fig_cs_comp}
\end{figure}

\begin{figure}[tbp!]
    \centering       \includegraphics[width=\linewidth,angle=0]{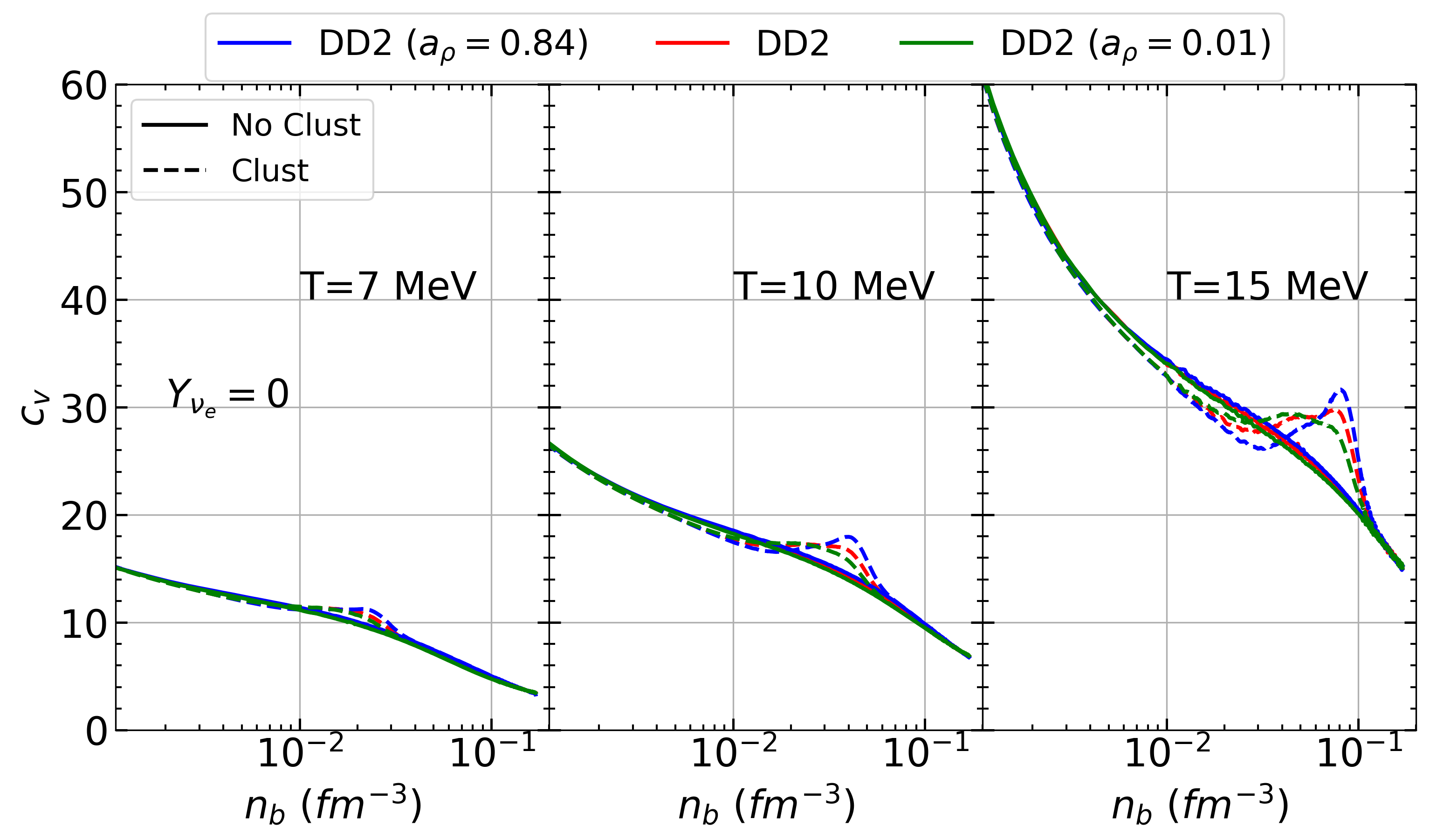}   
    \caption{Heat capacity per baryon, $c_V$, in neutrino-free $\beta$-equilibrated matter at $T=7$ MeV (left), $T=10$ MeV (middle), and $T=15$ MeV (right). Results obtained with (dashed lines) and without (solid lines) light clusters are shown for three DD2-based parametrizations characterized by $a_{\rho}=0.84$ (blue), $0.52$ (red), and $0.01$ (green).}
\label{Fig_cv_comp}
\end{figure}
Indeed, as shown in Fig.~\ref{Fig_yc_comp}, the cluster fractions increase as $L_{\rm sym}$ decreases, i.e., for increasing values of the symmetry energy, 
particularly around the densities at which
their abundances reach their maxima. The differences among the parametrizations are more pronounced for the $Z=2$ species, namely helions and $\alpha$ particles, owing to their stronger sensitivity to the available proton content. At fixed $Z$, however, the sensitivity to the symmetry energy slightly decreases with increasing neutron number $N$. These results demonstrate that even moderate variations in the total proton fraction can lead to appreciable changes in the light-cluster abundances.

A similar conclusion can be drawn for the isothermal squared speed of sound and the heat capacity per baryon, shown in Figs.~\ref{Fig_cs_comp} and~\ref{Fig_cv_comp}, respectively. The quantitative magnitude of the cluster-induced structures depends on the density dependence of the symmetry energy, through its influence on the equilibrium proton fraction and, consequently, on the cluster abundances. Nevertheless, the three parametrizations display the same qualitative behavior: light-cluster formation produces a visible structure in the isothermal squared speed of sound and a local enhancement of the heat capacity per baryon in the density region where clusters are most abundant. We conclude that, although their magnitude depends on the adopted isovector parametrization, the qualitative signatures identified in the previous subsections remain robust against the variations of the symmetry-energy density dependence explored here. Moreover, these features become more pronounced at the highest temperature considered, motivating future investigations at even higher temperatures. 

\section{Conclusions}
\label{sec:conclusions}
In this work, we have investigated the composition and thermodynamic
response of warm magnetized stellar matter within a gRMF framework including light nuclear clusters as
explicit quasiparticle degrees of freedom. We have considered deuterons,
tritons, helions, and \(\alpha\) particles, with in-medium dissolution
described through phenomenological binding-energy shifts. The effect of
strong magnetic fields has been incorporated through Landau quantization
of charged particles. We have analyzed both neutrino-free \(\beta\)-equilibrated matter and neutrino-trapped matter at fixed lepton
fraction, focusing on temperatures and densities relevant for warm
sub-saturation stellar matter.

A central result of our analysis is the key role played by the total proton content of the system, which also strongly controls light-cluster formation. Any mechanism that increases the proton fraction, such as neutrino trapping or the magnetic-field-induced rearrangement of the electron phase space, tends to favor the formation of proton-containing clusters. At the same time, the inclusion of light clusters feeds back on the equilibrium composition, leading to a larger total proton fraction in the density region where clusters are most abundant.  This highlights an interesting interplay between magnetic fields
and cluster formation: the magnetic field modifies the charged component of matter, which affects cluster formation, while clusters in turn rearrange the equilibrium charge
content. 

The role of neutrino trapping has been shown to be especially important.
At fixed lepton fraction, the larger proton content already present at
\(B=0\) enhances cluster formation and makes the  magnetic-field effect less pronounced.

The presence of clusters also strongly depletes the free-neutron component,
which can be significantly reduced in the region of maximal cluster abundance, motivating future extensions of our studies 
to lower temperatures, where
neutron superfluidity becomes relevant. Moreover, 
this modification of the free-neutron fraction could play a non-trivial role in determining the dynamical properties of the (proto-)NS inner crust, with  possible implications for phenomena such as glitches and oscillation modes.In addition, the redistribution of baryons between
free nucleons and light clusters may influence neutrino opacities and
charged-current reaction rates. 

Light clusters and magnetic fields also produce visible effects on the EOS. Clusters lower the energy at low densities, while they increase it in the density region where their abundances
approach their maxima, producing a corresponding stiffening of the pressure
followed by a softening when clusters start to dissolve. As temperature increases, these cluster-induced structures are shifted toward higher densities and remain visible over a broad density interval. The magnetic field, instead, mainly affects the very low-density region, where Landau
quantization modifies the charged-particle phase space and the equilibrium
composition, with an effect that becomes weaker and more localized as temperature increases.

These modifications of the EOS leave clear imprints on the thermodynamic
response functions. In particular, the cluster-induced change in the
stiffness of the EOS produces characteristic structures in the isothermal squared speed of sound. The signature is more pronounced in neutrino-trapped matter, where the larger cluster fractions
generate a clear local maximum and minimum associated with cluster
formation and dissolution. A similar effect is found in the heat capacity per baryon:
light clusters produce a local enhancement in the density region where
their abundances are maximal. This highlights the importance of a more complete, composition-dependent description of warm stellar matter for
predicting not only its  equation of state, but also its mechanical and thermal response.

Finally, the comparison among three RMF parametrizations, constructed by varying the isovector sector while keeping the isoscalar sector and $E_{\rm sym}(n_{\rm sat})$ unchanged, shows that the magnitude of the cluster-induced effects depends on the subsaturation density dependence of the symmetry energy. This dependence arises from the influence of the symmetry energy on
the equilibrium proton fraction and, consequently, on the cluster abundances. Nevertheless, the qualitative signatures in the EOS, the isothermal squared speed of sound and the heat capacity per baryon, remain robust over the range of isovector behaviors explored here.

This work therefore provides a microscopic baseline for understanding how magnetic fields, weak-equilibrium conditions, and light-cluster formation jointly affect the composition and thermodynamic response of warm sub-saturation stellar matter. More generally, our findings highlight the
importance of including composition-dependent thermal effects in
finite-temperature EOSs aimed at interpreting proto-NS, merger remnants, and future multimessenger observations.

\section{Acknowledgements}
Stimulating discussions with Stefan Typel, Helena Pais and Micaela Oertel are gratefully acknowledged.

\bibliographystyle{unsrt}
\bibliography{ref}

\end{document}